\documentclass[oneside, onecolumn, draftcls, peerreview, 12pt]{IEEEtran}

\newcommand{\sss}{\scriptscriptstyle}
\newcommand{\mbs}[1]{\boldsymbol{#1}}
\renewcommand{\mbs}[1]{\mathbf{#1}}
\renewcommand{\mbs}[1]{\pmb{#1}}

\newcommand{\vect}[1]{{\lowercase{\mbs{#1}}}}
\newcommand{\mat}[1]{{\uppercase{\mbs{#1}}}}

\newcommand{\asympteq}{\doteq} %

\usepackage{footnote} 
\makesavenoteenv{tabular}
\usepackage{color}
\usepackage{url}
\usepackage{cite}
\usepackage{graphicx}
\usepackage{graphics}
\usepackage{psfrag}
\usepackage{array}
\usepackage{mdwtab}
\usepackage{epsfig}
\usepackage{subfigure}
\usepackage{amsmath,amssymb,amsfonts,mathrsfs,amscd}
\usepackage{pst-node,pst-text,pst-3d}
\usepackage{pstricks}

\newtheorem{definition}{Definition}
\newtheorem{theorem}{Theorem}

\newtheorem{corollary}{Corollary}

\newtheorem{lemma}{Lemma}

\newcommand{\bb}[1]{\mathbb{#1}}
\newcommand{\nnb}{\nonumber}

\newcommand{\Fig}[1]{Fig.~\!\ref{#1}}

\newcommand{\Eq}[1]{(\ref{#1})}

\newcommand{\ie}{\emph{i.e.}}
\newcommand{\eg}{\emph{e.g.}}

\newcommand{\etal}{\emph{et al.}\,}

\newcommand{\D}{\displaystyle}

\newcommand{\eqncaseslabel}[6]{
  \begin{equation}
    \setlength{\nulldelimiterspace}{0pt}
    #1=\left\{
      \begin{IEEEeqnarraybox}[\relax][c]{l's}
        #2, &for #3\\
        #4, &for #5%
      \end{IEEEeqnarraybox}\right.\label{#6}
  \end{equation}
}

\newcommand{\eqncasesasympt}[5]{
  \begin{equation}
    \setlength{\nulldelimiterspace}{0pt}
    #1\asympteq\left\{
      \begin{IEEEeqnarraybox}[\relax][c]{l's}
        #2, &#3\\
        #4, &#5%
      \end{IEEEeqnarraybox}\right.
  \end{equation}
}

\newcommand{\eqncasesasymptlabel}[6]{
  \begin{equation}
    \setlength{\nulldelimiterspace}{0pt}
    #1\asympteq\left\{
      \begin{IEEEeqnarraybox}[\relax][c]{l's}
        #2, &#3\\
        #4, &#5%
      \end{IEEEeqnarraybox}\right.\label{#6}
  \end{equation}
}

\renewcommand{\matrix}[1]{\begin{bmatrix}#1\end{bmatrix}}

\newcommand\transcsymbol{\scriptscriptstyle \dag \!}
\newcommand\transsymbol{\scriptscriptstyle \mathsf{T} \!}

\newcommand\Abs[1]{\left|#1\right|}

\newcommand{\transc}[1]{{#1}^{\transcsymbol}}
\newcommand{\trans}[1]{{#1}^{\transsymbol}}
\newcommand{\pstv}[1]{{#1}^{\sss +}}

\newcommand\diag{\mathrm{diag}}

\newcommand\Det{\mathrm{Det}}
\newcommand\Norm[1]{\left\|{#1}\right\|}

\newcommand\defeq{\triangleq}

\newcommand{\Id}{\mathbf{I}}
\newcommand{\CN}[1][\Id]{\Ccal\Ncal\!\left(0,#1\right)}

\newcommand{\SNR}{{\mathsf{SNR}} }

\newcommand\Pout{P_{\textrm{out}}}

\renewcommand\d{\mathrm{d}}

\newcommand\CC{\bb{C}}
\newcommand\EE{\bb{E}}

\newcommand\RR{\bb{R}}

\newcommand\Ccal{\mathcal{C}}

\newcommand\Ical{\mathcal{I}}

\newcommand\Ncal{\mathcal{N}}

\newcommand\Rcal{\mathcal{R}}
\newcommand\Scal{\mathcal{S}}

\newcommand\Wcal{\mathcal{W}}

\newcommand{\mA}{\mat{A}}

\newcommand{\mD}{\mat{D}}

\newcommand{\mG}{\mat{G}}
\newcommand{\mH}{\mat{H}}

\newcommand{\mM}{\mat{M}}

\newcommand{\mP}{\mat{P}}
\newcommand{\mQ}{\mat{Q}}
\newcommand{\mR}{\mat{R}}
\newcommand{\mS}{\mat{S}}
\newcommand{\mT}{\mat{T}}

\newcommand{\mW}{\mat{W}}

\newcommand{\mc}{\vect{c}}

\newcommand{\mm}{\vect{m}}
\newcommand{\mn}{\vect{n}}

\newcommand{\mx}{\vect{x}}
\newcommand{\my}{\vect{y}}
\newcommand{\mz}{\vect{z}}

\newcommand{\mXi}{\mbs{\Xi}}

\newcommand{\mPi}{\mat{\Pi}}

\newcommand{\mPhi}{\mbs{\Phi}}

\newcommand{\mlambda}{\boldsymbol{\lambda}}
\newcommand{\malpha}{\boldsymbol{\alpha}}

\newcommand{\mmu}{\boldsymbol{\mu}}
\newcommand{\mbeta}{\boldsymbol{\beta}}

\usepackage{enumerate}

\usepackage{./cases}

\begin{document}
\title{Diversity of MIMO Multihop Relay Channels---Part~I: Amplify-and-Forward}
 \author{%
   \authorblockN{Sheng Yang and Jean-Claude Belfiore\thanks{Manuscript
       submitted to the IEEE Transactions on Information Theory.  The
       authors are with the Department of Communications and
       Electronics, \'{E}cole Nationale Sup\'{e}rieure des
       T\'{e}l\'{e}communications, 46, rue Barrault,
    75013 Paris, France~(e-mail: syang@enst.fr; belfiore@enst.fr).}}}

\maketitle

\newcommand{\nT}{n_\text{T}}
\newcommand{\nR}{n_\text{R}}
\newcommand{\nS}{n_\text{S}}
\newcommand{\PhiT}{\mPhi_\text{T}}
\newcommand{\PhiR}{\mPhi_\text{R}}
\newcommand{\PhiS}{\mPhi_\text{S}}
\newcommand{\asymptleq}{\ \dot{\leq}\,}
\newcommand{\asymptgeq}{\ \dot{\geq}\,}
\newcommand{\mnt}{\mbs{\tilde{n}}}
\newcommand{\E}{E}
\newcommand{\Ec}{\hat{E}}
\newcommand{\ilb}{\underline{i}}
\newcommand{\jlb}{\underline{j}}
\newcommand{\wrt}{\emph{w.r.t.}}
\newcommand{\node}[1]{node~\#$#1$}

\newcommand{\nminp}{n'_{\min}}
\newcommand{\nmin}{n_{\min}}
\newcommand{\DMT}{\textsf{DMT}}
\renewcommand{\div}{\textsf{div}}

\newcommand{\dAF}{d^{\text{AF}}}
\newcommand{\dDF}{d^{\text{DF}}}
\newcommand{\dPF}{d^{\text{PF}}}
\newcommand{\RP}{\mPi}

\newcommand{\Ndc}{N_{\text{d/c}}}
\newcommand{\Ndcs}{N^*_{\text{d/c}}}

\begin{abstract}
  In this two-part paper, we consider the multiantenna multihop relay
  channels in which the source signal arrives at the destination
  through $N$ independent relaying hops in series. The main concern of
  this work is to design relaying strategies that utilize efficiently
  the relays in such a way that the diversity is maximized. In part~I,
  we focus on the amplify-and-forward~(AF) strategy with which the
  relays simply scale the received signal and retransmit it. More
  specifically, we characterize the diversity-multiplexing
  tradeoff~(DMT) of the AF scheme in a general multihop channel with
  arbitrary number of antennas and arbitrary number of hops. The DMT
  is in closed-form expression as a function of the number of antennas
  at each node. First, we provide some basic results on the DMT of the
  general Rayleigh product channels. It turns out that these
  results have very simple and intuitive interpretation. Then,
  the results are applied to the AF multihop channels which is shown
  to be equivalent to the Rayleigh product channel, in the DMT sense. Finally, the
  project-and-forward~(PF) scheme, a variant of the AF scheme, is
  proposed. We show that the PF scheme has the same DMT as the AF
  scheme, while the PF can have significant power gain over the AF
  scheme in some cases. In part II, we will derive the upper bound on
  the diversity of the multihop channels and show that it can be
  achieved by partitioning the multihop channel into AF subchannels.  
\end{abstract}


\IEEEpeerreviewmaketitle

\section{Introduction and Problem Description}
\label{sec:intro}
Wireless relaying systems have lots of advantages over traditional
direct transmission systems. For example, the periphery can be
extended by the relays and the coverage of the existing network can be
improved. Using relays can also shorten the point to point
transmission distance, which results in lower power~(interference)
level or in higher throughput. Furthermore, all these benefits can be
realized in a more flexible, easier and cheaper to deploy network.

Recently, there has been a boosting interest in the cooperative
diversity with which the spatial diversity is exploited through
distributed relays. Since the work of Sendonaris
\etal~\cite{Sendonaris1,Sendonaris2} that introduced the notion of
cooperative diversity, a number of relaying protocols have been
proposed~(see, \eg,
\cite{LTW1,LTW2,Nabar,Jing,ElGamal_coop,SY_JCB_coop,SY_JCB_SAF,Elia_relay}).
Most of the previous works consider the single-antenna two-hop relay
channel where the source signal is able to arrive at the destination
through at most two hops, \ie, the source-relay hop and
relay-destination hop. In an $N$-relay channel, it is shown that a
diversity order of $N+1$~(respectively, $N$) can be achieved
with~(respectively, without) the direct source-destination link.

In this work, we consider the MIMO multihop channel model without
direct source-destination link. That is, the source signal arrives at
the destination through $N$ independent relaying hops in series. In the
two-hop case, our model is reduced to the model studied by Jing and
Hassibi~\cite{Jing}.
The central concern of our work is to design relaying strategies that
utilize efficiently the relays in such a way that the diversity is
maximized. In part~I, we focus on the amplify-and-forward~(AF)
strategy with which the relays simply scale the received signal and
retransmit it. The main contributions of this paper are as follows.
\begin{enumerate}
\item First, we obtain the diversity-multiplexing tradeoff~(DMT) of
  the Rayleigh product channel, whose channel matrix is a product of
  independent Gaussian matrices. It turns out that each Rayleigh
  product channel belongs to an equivalent class that is uniquely
  represented by the so-called \emph{minimal form}. Furthermore, based
  on the closed-form expression of the DMT, we derive a recursive DMT
  characterization that have very simple and intuitive interpretation.
\item Then, it is shown that the AF multihop channel is actually
  equivalent to the Rayleigh product channel. We can thus identify the
  two channels and all previously established results apply to the
  multihop channel. Therefore, the diversity properties of the AF
  multihop channel in terms of the number of hops and the number of
  antennas in each node are completely characterized. We also propose
  the project-and-forward~(PF) scheme, a variant of the AF scheme, in
  the case where full antenna cooperation is possible. It is shown
  that, although the PF scheme has the same DMT as the AF scheme, the
  PF can have significant power gain over the AF scheme in some cases.
\item Finally, it is pointed out that using less relaying antennas
  improve the power gain by avoiding the \emph{hardening} of relayed noise, a
  particular phenomenon in the AF multihop channel. And reducing the
  number of transmit antennas can lower significantly the coding delay
  and decoding complexity. The vertical channel reduction result gives
  exactly the minimum number of antennas we need at each node to keep
  the same DMT.
\end{enumerate}

In part~II of this paper, we will derive an upper bound on the
diversity of the multihop channels and show that the AF scheme is not
optimal in general. Then, we will proposed both distributed and
non-distributed schemes that achieve the upper bound. The main idea is
to partition the multihop channel into AF subchannels.

The rest of part~I is organized as follows.
Section~\ref{sec:system-model} presents the channel model and the AF
scheme with some basic assumptions. The Rayleigh product channel is
introduced and studied in section~\ref{sec:rp}. Results concerning the
AF and PF schemes are collected in section~\ref{sec:af}. In
section~\ref{sec:nr}, numerical results on some typical scenarios are shown. Finally, we
draw a brief conclusion in section~\ref{sec:conclusion}. For
fluidity of the presentation, all demonstrations of proofs are delayed
to the appendices.

In this paper, we use boldface lower case letters $\mbs{v}$ to denote
vectors, boldface capital letters $\mbs{M}$ to denote matrices.
$\Ccal\Ncal$ represents the complex Gaussian random variable.
$\trans{[\cdot]},\transc{[\cdot]}$ respectively denote the matrix
transposition and conjugated transposition operations. $\Norm{\cdot}$
is the vector norm. $\pstv{(x)}$ means $\max(0,x)$. $\Det(\mM)$ is the
absolute value of the determinant $\det(\mM)$. The square root
$\sqrt{\mP}$ of a positive semi-definite matrix $\mP$ is defined as a
positive semi-definite matrix such that
$\mP=\sqrt{\mP}\transc{\bigl(\sqrt{\mP}\bigr)}$. The ordered
eigenvalues of a positive semi-definite matrix $\mP$ are denoted by
$\mlambda(\mP)$ or $\mmu(\mP)$. We define $\malpha(\mP)$ and
$\mbeta(\mP)$ by
$$\alpha_i(\mP )\defeq -\log \lambda_i(\mP)/\log\SNR
\quad\text{and}\quad \beta_i(\mP )\defeq -\log \mu_i(\mP)/\log\SNR.$$
And we call them the \emph{eigen-exponents} of $\mP$, with a slight abuse of
terminology. We drop the arguments of $\mlambda,\mmu,\malpha,\mbeta$
when confusion is not likely. For any quantity $q$,
  \begin{equation*}
    q \asympteq \SNR^{a}\quad \textrm{means}\quad \lim_{\SNR\to\infty}\frac{\log q}{\log \SNR} = a
  \end{equation*}
  and similarly for $\asymptleq$ and $\asymptgeq$.  The tilde notation
  $\tilde{\mn}$ is used to denote the (increasing)~ordered version of
  $\mn$. Let $\mm$ and $\mn$ be two vectors of same length $L$, then
  $\mm\preceq\mn$ means $\tilde{m}_i\leq\tilde{n}_i$, $\forall\,i$.

\section{System Model}
\label{sec:system-model}

\subsection{Channel Model}
\label{sec:channel-model}

\begin{figure*}[!t]
  \begin{center}
         \includegraphics[angle=0,width=0.5\textwidth]{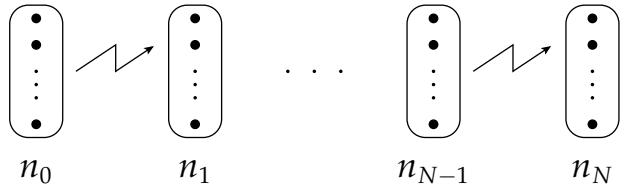}
         \caption{A MIMO multihop relay channel.}
         \label{fig:multihop}    
  \end{center}
\end{figure*}

The considered $N$-hop relay channel model is illustrated in
\Fig{fig:multihop}, where there are one source~(\node{0}), one
destination~(\node{N}), and $N-1$ clusters of intermediate relays. Each
cluster is logically seen as a node~(\node{1} to \node{N-1}) that is
equipped with multiple antennas~($n_i$ antennas for \node{i}). We
assume that \node{i} can only hear \node{i-1}. Mathematically, we have 
\begin{equation*}
  \my_{i} = \mH_{i} \mx_{i-1} + \mz_{i} 
\end{equation*}
where $\mH_i\in\CC^{n_i\times n_{i-1}}$ is the channel between
\node{i-1} and \node{i}; $\mx_i,\my_i\in\CC^{n_i\times1}$ is the
transmitted and received signal at \node{i};
$\mz\in\CC^{n_i\times1}\in\CN$ is the additive white Gaussian noise at
\node{i}. The channels $\mH_i$'s are independent and modeled as
Rayleigh quasi-static channels, \ie, the entries of $\mH_i$ are i.i.d.
$\CN[1]$ distributed and do not change during the transmission of a
data frame. For simplicity, it is assumed that the intermediate nodes
work in full-duplex\footnote{The assumption is merely for simplicity
  of notation. As one can easily verify, since no cross-talk between
  different channels, the half-duplex constraint is directly
  translated to a reduction of degrees of freedom by a factor of two
  and does not impact the relaying strategy. This is achieved by
  letting all even-numbered~(respectively, odd-numbered) nodes
  transmit~(respective, receive) in even-numbered time slot and
  received~(respective, transmit) in odd-numbered time slots.} mode
and all transmitting nodes are subject to the same short-term power
constraint
\begin{equation}
  \label{eq:power-constraint}
  \EE\{\Norm{\mx_i}^2\} \leq \SNR,\quad\forall\,i
\end{equation}%
where the expectation is taken on the noises. All terminals are
supposed to have full channel state information~(CSI) at the receiver
and no CSI at the transmitter. From now on, we denote the channel as a
$(n_0,n_1,\ldots,n_N)$ multihop channel.

\subsection{Amplify-and-Forward Protocol}
\begin{figure}
\begin{center}
\includegraphics[height=0.15\textwidth]{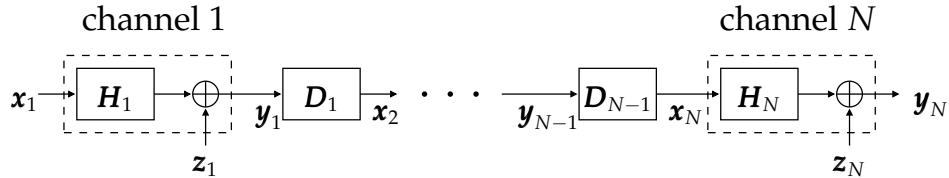}
\caption{Amplify-and-forward strategy for multihop channels.}
\label{fig:multihop_AF_DF}
\end{center}
\end{figure}

The AF strategy is described as follows. At each node, the received
signal of each antenna is normalized to the same power level and then
retransmitted. As shown in Fig.~\ref{fig:multihop_AF_DF}, the signal model is
\begin{align*}
  \my_i &= \mH_i \mx_i + \mz_i, \\
  \mx_{i+1} &= \mD_{i} \my_{i}
\end{align*}%
where the transmitted signal $\mx_i$ has the short-term power
constraint
$$\EE\left(\Abs{\mx_i[j]}^2\right) \leq \frac{\SNR}{n_i};$$
the
scaling matrix $\mD_i\in\CC^{n_i\times n_i}$ is diagonal with the
normalization factors\footnote{In the case where long-term power
  constraint is imposed, we simply replace the channel coefficients
  $\Abs{\mH_i[j,k]}$ in \Eq{eq:mD} by $1$'s.}
\begin{equation}
  \label{eq:mD}
\mD_i[j,j] =
\sqrt{\frac{1}{\frac{\SNR}{n_{i-1}}\left(\sum_{k=1}^{n_{i-1}}\Abs{\mH_i[j,k]}^2\right)+1}}\cdot
\sqrt{\frac{\SNR}{n_{i}}}.  
\end{equation}

\subsection{Diversity-Multiplexing Tradeoff}

In this paper, we use the diversity-multiplexing tradeoff~(DMT) as the
performance measure.
\begin{definition}[Multiplexing and diversity gains\cite{Zheng_Tse}]
  The \emph{multiplexing gain} $r$ and \emph{diversity gain} $d$ of a
  fading channel are defined by
\begin{equation*}
  r \defeq \lim_{{\sss\textsf{SNR}}\to\infty} \frac{R(\SNR)}{\log\SNR} \quad
\textrm{and}\quad 
  d \defeq - \lim_{{\sss\textsf{SNR}}\to\infty} \frac{\log\Pout(\SNR,R)}{\log\SNR}
\end{equation*}
where $R(\SNR)$ is the target data rate and $\Pout(\SNR,R)$ is the
outage probability for a target rate $R$. A more compact form is
\begin{equation}
  \label{eq:dmt-compact}
  \Pout(\SNR,r\log\SNR) \asympteq \SNR^{-d}.
\end{equation}%
\end{definition}
Note that in the definition we use the outage probability instead of
the error probability, since it is shown in \cite{Zheng_Tse} that the
error probability is dominated by the outage probability in the high
SNR regime and that the thus defined DMT is the best that we can
achieve with any coding scheme.

\begin{lemma}\label{lemma:dmt-Rayleigh}
  The DMT of a $n_\text{t}\times n_\text{r}$ Rayleigh channel is a
  piecewise-linear function connecting the points~$(k,d(k))$,
  $k=0,1,\ldots,\min{(n_\text{t},n_\text{r})}$, where
  \begin{equation*}
    d(k) = (n_\text{t}-k)(n_\text{r}-k).
  \end{equation*}%
\end{lemma}

\section{The Rayleigh Product Channel}
\label{sec:rp}

As it is shown in the next section, the AF multihop channels are
intimately related to a more general Rayleigh product channel defined
below. In this section, we investigate the Rayleigh product channel
and provides some basic results on the diversity. Let us begin by the following definitions. 
\begin{definition}[Rayleigh product channel]
  Let $\mH_i\in\CC^{n_{i-1}\times n_i}$, $i=1,2,\ldots,N$, be $N$
  independent complex Gaussian matrices with i.i.d. zero mean unit
  variance entries. A $(n_0,n_1,\ldots,n_N)$ Rayleigh product channel
  is a $n_N\times n_0$ MIMO channel defined by
  \begin{equation}
    \label{eq:channel-model}
    \my = \sqrt{\frac{\SNR}{n_1\cdots n_{N}}} \RP\,\mx + \mz
  \end{equation}%
  where $\RP \defeq \mH_1\mH_2\cdots\mH_N$; $\mx$ is the transmitted
  signal with power constraint~$\EE(\Norm{\mx}^2) \leq n_N$;
  $\mz\in\CC^{n_0\times1}\sim\CN$ is the additive white Gaussian
  noise; $\SNR$ is the receive signal-to-noise ratio~(SNR) per
  receive antenna.
\end{definition}

\begin{definition}[Exponential equivalence]
  Two channels are said to be \emph{exponentially equivalent} or
  \emph{equivalent} if their eigen-exponents have the same
  asymptotical joint pdf.
\end{definition}

Let $\mnt$ be the ordered version of $\mn$ with $\tilde{n}_N\geq
  \tilde{n}_{N-1}\geq\cdots\geq \tilde{n}_0$.
\begin{definition}[Reduction of Rayleigh product channel]
  A $(m_0,m_1,\ldots,m_k)$ Rayleigh product channel is said to be a
  \emph{reduction} of a $(n_0,n_1,\ldots,n_N)$ Rayleigh product
  channel if 1)~they are equivalent, 2)~$k\leq N$, and
  3)~$(m_{0}, m_{1}, \ldots, m_{k}) \preceq
  (\tilde{n}_{0}, \tilde{n}_{1}, \ldots, \tilde{n}_{k})$. In particular, if $k=N$, then it is called a
  \emph{vertical reduction}.  Similarly, if $\tilde{m}_i=\tilde{n}_i,\ 
  \forall\,i\in[0,k]$, it is a \emph{horizontal reduction}.
\end{definition}

\begin{definition}[Minimal form]
  $(\tilde{n}_0,\tilde{n}_1,\ldots,\tilde{n}_{N^*})$ is said to be a
  \emph{minimal form} if no reduction other than itself exists.
  Similarly, it is called a \emph{minimal vertical
    form}~(respectively, \emph{minimal horizontal form}) if no
  vertical~(respectively, horizontal) reduction other than itself
  exists. A channel is said to have \emph{order} $N^*$ if its minimal
  form is of length $N^*+1$.
\end{definition}%

\subsection{Joint PDF of the Eigen-exponents of $\RP\transc{\RP}$}
\label{sec:joint-pdf}

\begin{theorem}\label{thm:asympt-pdf}
  Let us denote the non-zero ordered eigenvalues of
  $\RP\transc{\RP}$ by $\lambda_1\geq\cdots\geq\lambda_{\nmin}>0$
  with ${\nmin}\defeq\D\min_{i=0,\ldots,N} n_i$. 
  Then, the joint pdf of the eigen-exponents $\malpha$ satisfies
  \eqncasesasymptlabel{p(\malpha)}{\SNR^{-\E(\malpha)}}{for
    $0\leq\alpha_1\leq\ldots\leq\alpha_{\nmin}$,}{\SNR^{-\infty}}{otherwise}{eq:p-malpha}
  where 
  \begin{equation}
    \label{eq:Ea}
    \E(\malpha) \defeq \sum_{i=1}^{\nmin}c_i\alpha_i
  \end{equation}%
  with 
  \begin{equation}\label{eq:ci}
    c_i \defeq 1-i + \min_{k=1,\ldots, N} \left\lfloor\frac{\sum_{l=0}^{k}\tilde{n}_l - i}{k} \right\rfloor,\quad i=1,\ldots,{\nmin}.
  \end{equation}%
\end{theorem}
By definition, ${\nmin} = \tilde{n}_0$ and we interchange the notations
depending on the context. From the theorem, we can see that the
asymptotical eigen-exponents distribution depends only on
$(\tilde{n}_0,\tilde{n}_1,\ldots,\tilde{n}_N)$, the ordered version of
$({n}_0,{n}_1,\ldots,{n}_N)$. For example, a $(3,1,4,2)$ channel is
equivalent to a $(1,2,3,4)$ channel, in the eigen-exponent sense. 

\begin{theorem}\label{thm:reduction}
  A $(n_0,n_1,\ldots,n_N)$ Rayleigh product channel can be reduced to a $(\tilde{n}_0,\tilde{n}_1,\ldots,\tilde{n}_k)$ channel if and only if
  \begin{equation}
    \label{eq:reduction-cond}
    k(\tilde{n}_{k+1} + 1) \geq \sum_{l=0}^k \tilde{n}_l.    
  \end{equation}%
  In particular, it can be reduced to a Rayleigh channel if and only if
  \begin{equation}
    \label{eq:reduction-cond-0}
    \tilde{n}_{2} + 1 \geq \tilde{n}_0+\tilde{n}_1.    
  \end{equation}%
\end{theorem}
This theorem implies that $(\tilde{n}_0,\tilde{n}_1,\ldots,\tilde{n}_{N^*})$ is a minimal form if there exists no $k<N^*$ such that \Eq{eq:reduction-cond} is satisfied.
One can also verify that if
$(\tilde{n}_0,\tilde{n}_1,\ldots,\tilde{n}_{N^*})$ is a minimal
horizontal form of $(n_0,n_1,\ldots,n_N)$, then 1)~it is also a
minimal form; and 2)~the minimal vertical form is
$(\tilde{n}_0,\tilde{n}_1,\ldots,\tilde{n}_{N^*},\bar{n},\ldots,\bar{n})$
where
\begin{equation}
  \label{eq:minimum_number}
  \bar{n} = \left\lceil \frac{\sum_{l=0}^{N^*} \tilde{n}_i}{N^*} - 1\right\rceil.  
\end{equation}
Furthermore, note that the order $N^*$ is upper-bounded by
$\tilde{n}_0$ because \Eq{eq:reduction-cond} is always satisfied with
$k=\tilde{n}_0$. In other words, the length of the minimal form is
bounded by $\tilde{n}_0+1$. In particular, the minimal form of a
$(1,n_1,\ldots,n_N)$ Rayleigh product channel is always $(1,n_1)$,
\ie, a $1 \times \tilde{n}_1$ or $\tilde{n}_1 \times 1$ Rayleigh
channel.

\begin{theorem}\label{thm:minimal}
  Two Rayleigh product channels are equivalent if and only if they
  have the same minimal form.
\end{theorem}
From this theorem, we deduce that the class of exponential equivalence is
\emph{uniquely} identified by the minimal form. Therefore, $N^*$ can also be
defined as the order of the class.

\subsection{Characterization of the Diversity-Multiplexing Tradeoff}

From theorem~\ref{thm:asympt-pdf}, we can derive the
DMT of a Rayleigh product channel.
\begin{theorem}[Direct characterization]\label{thm:dmt-rp}
  The DMT of a Rayleigh product channel~$(n_0,n_1,\ldots,n_N)$ is a
  piecewise-linear function connecting the points~$(k,d(k))$,
  $k=0,1,\ldots,\nmin$, where
  \begin{equation}
    d(k) = \sum_{i=k+1}^{\nmin} c_i\label{eq:dk}
  \end{equation}%
  with $c_i$ defined by \Eq{eq:ci}. 
\end{theorem}
Since the DMT is a bijection of the coefficients $c_i$'s, all results
obtained previously apply to the DMT and two Rayleigh product channels
are equivalent if and only if they have the same DMT. Hence, the
exponential equivalence class is also the DMT-equivalence class.
However, unlike the eigen-exponent, the DMT provides an insight on the
diversity performance of a channel~(or a scheme) for different
multiplexing gain. Note that, despite the closed-form nature of the
characterization \Eq{eq:dk}, it is lack of intuition. That is why we
search for an alternative characterization.  \newcommand{\R}{R}
\begin{figure*}[!t]
\begin{center}
  \subfigure[Interpretation of
  $\R_1^{(N)}(k)$]{\includegraphics[angle=0,height=0.3\textwidth]{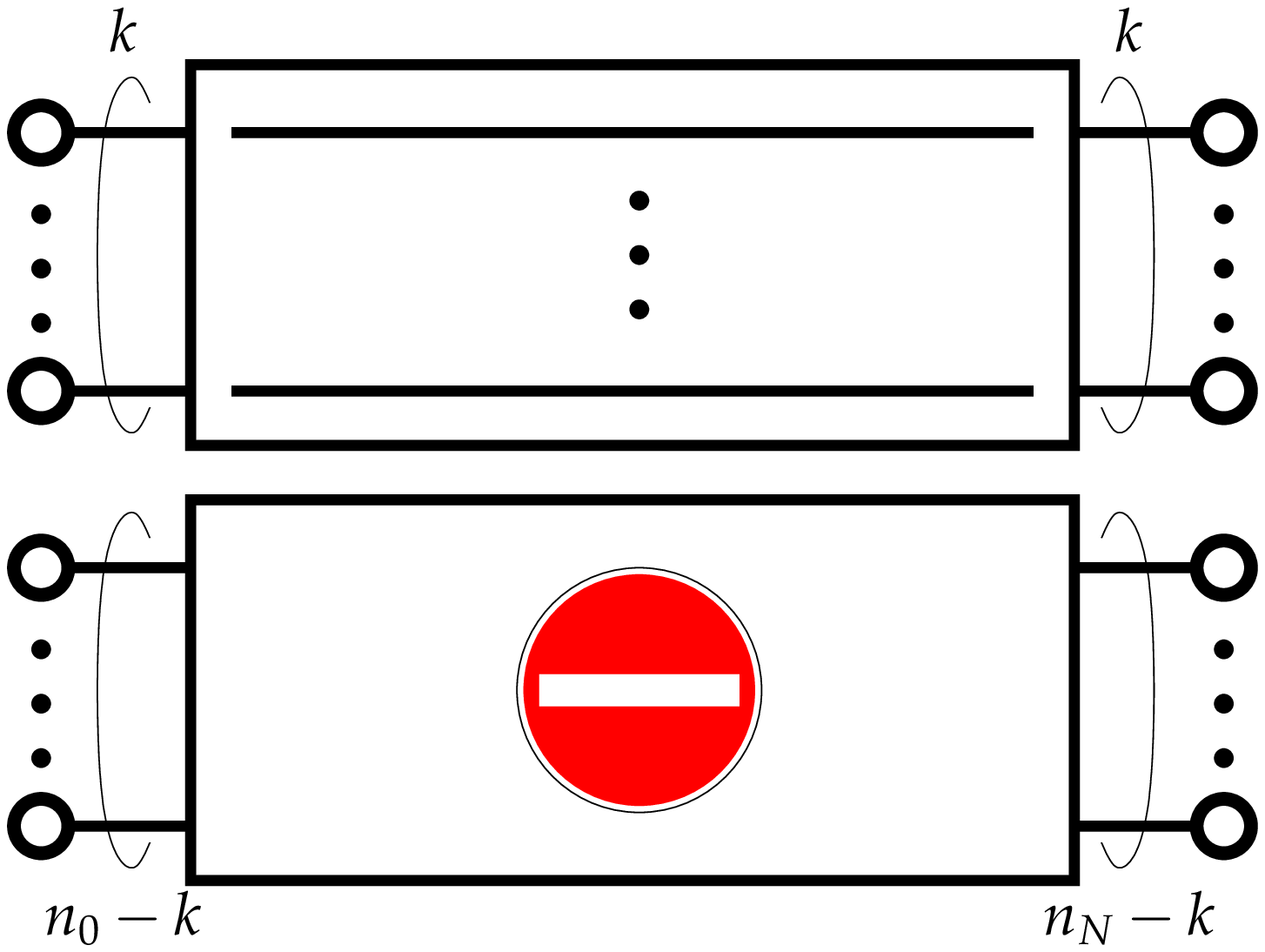}\label{fig:interp1}}
  \hspace{2cm} \subfigure[Interpretation of
  $\R_2^{(N)}(i)$]{\includegraphics[angle=0,height=0.3\textwidth]{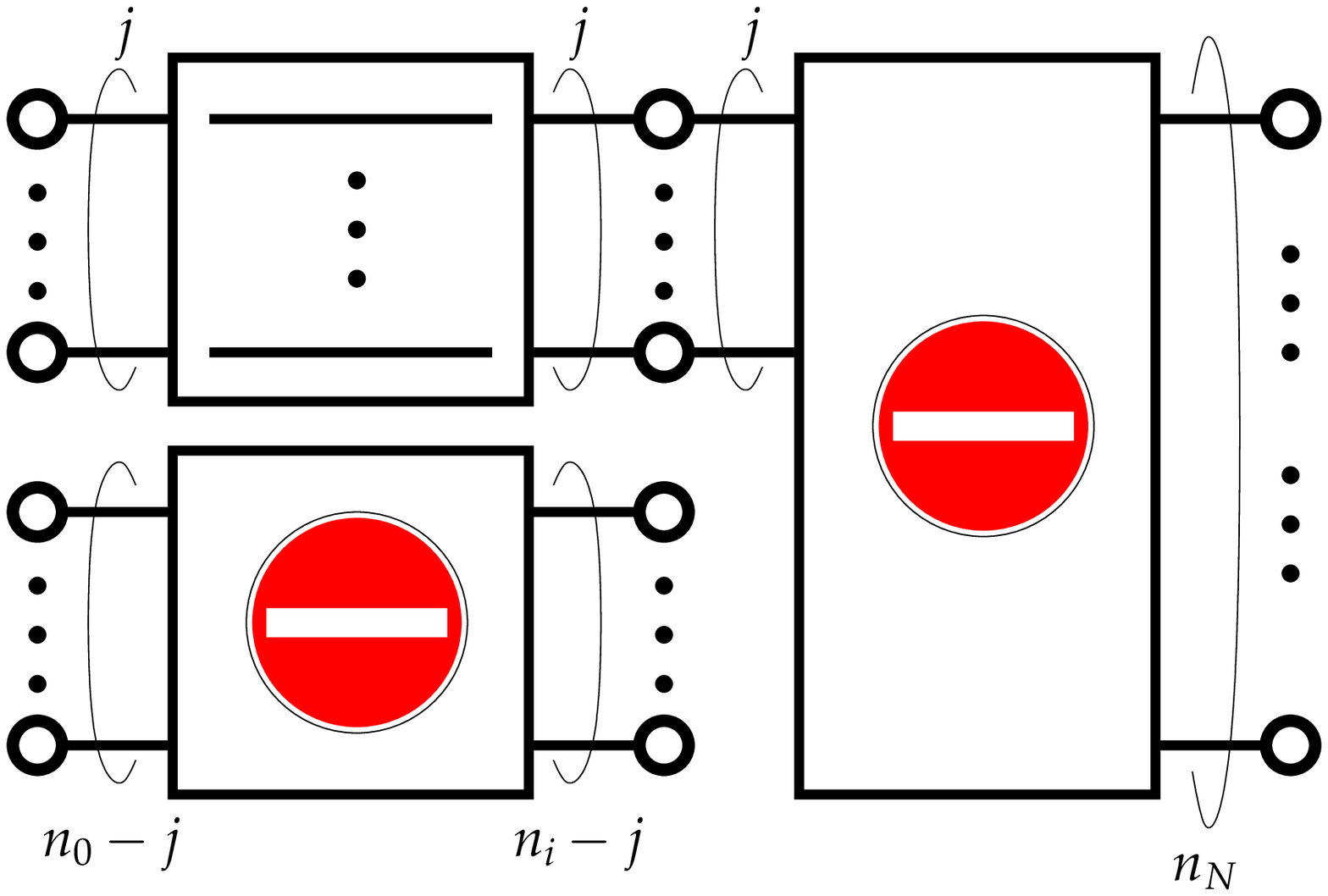}\label{fig:interp2}}
\caption{Interpretations of the DMT of the Rayleigh product channel.}
\label{fig:interp}  
\end{center}
\end{figure*}
\begin{theorem}[Recursive characterization]\label{thm:recursive}
  The DMT $d(k)$ defined in \Eq{eq:dk} can be alternatively
  characterized by
  \begin{align} 
    \R_1^{(N)}(k)~:\quad d_{(n_0,\ldots,n_N)}(k) &= d_{(n_0-k,\ldots,n_N-k)}(0),\quad\forall k;\label{eq:interp1} \\
    \R_2^{(N)}(i)~:\quad d_{(n_0,\ldots,n_N)}(0) &= \min_{j\geq 0}
    d_{(n_0,\ldots,n_i)}(j) +
    d_{(j,n_{i+1},\ldots,n_N)}(0),\quad\forall i;
    \label{eq:interp2} \\
    \R_3^{(N)}(i,k)~:\quad d_{(n_0,\ldots,n_N)}(k) &= \min_{j\geq k}
    d_{(n_0,\ldots,n_i)}(j) +
    d_{(j,n_{i+1},\ldots,n_N)}(k),\quad\forall i,k.
    \label{eq:interp3} 
  \end{align}%
\end{theorem}%
The recursive characterization has an intuitive interpretation as
follows. Let us consider $k$ as a ``\emph{network flow}'' between the
source and the destination and $d(k)$ as the \emph{minimum ``cost''}
to limit the flow to $k$~(the flow-$k$ event).  In particular, the
maximum diversity $d(0)$ can be seen as the \emph{``disconnection
  cost''}. First, $\R_1(k)$ says that the most efficient way to limit
the flow to $k$ is to keep a $(k,k,\ldots,k)$ channel fully connected
and to disconnect the $(n_0-k,n_1-k,\ldots,n_N-k)$ residual channel,
as shown in \Fig{fig:interp1}. Then, $\R_2(i)$ suggests that in order
to disconnect a $(n_0,n_1,\ldots,n_N)$ channel, if we allow for $j$
flows from the source to some node $i$, then the
$(j,n_{i+1},\ldots,n_N)$ channel from the $j$ ``ends'' of the flows at
node~$i$ to the destination must be disconnected. The idea is shown in
\Fig{fig:interp2}.  Obviously, the most efficient way is such that the
total cost is minimized with respect to $j$. This interpretation sheds
lights on the typical outage event of the Rayleigh product channel. In
the trivial case of $N=1$ (the Rayleigh channel), there is only one
subchannel. The typical and only way for the channel to be in outage
is that all the paths are bad, \ie, the disconnection cost is
$\tilde{n}_0\times\tilde{n}_1$.  In the non-trivial cases, there are
more than one subchannels and thus the typical outage event is not
necessarily for one of the subchannels being totally bad. The
\emph{mismatch} of two partially bad subchannels can also cause
outage.  In a more general way, the flow-$k$ event takes place when
both the flow-$j$ event in the $(n_0,\ldots,n_i)$ channel and the
flow-$k$ event in the $(j,n_{i+1},\ldots,n_k)$ channel happen at the
same time.
We can verify that $(\R_1(k),\R_3(i,k))$ is equivalent to
$(\R_1(k),\R_2(i))$. Note that the DMT is completely characterized by
these relations in a recursive manner.

The following corollaries conclude some properties of the DMT of the
Rayleigh product channel. 

\begin{corollary}[Monotonicity]\label{coro:increase}
The DMT is monotonic in the following senses~:
  \begin{enumerate}
  \item if $(n_{1,0}, n_{1,1}, \ldots, n_{1,N}) \succeq ( n_{2,0},
    n_{2,1}, \ldots, n_{2,N} )$, 
    then
    $$d_{(n_{1,0}, \ldots, n_{1,N})}(r) \geq d_{(n_{2,0}, \ldots,
      n_{2,N})}(r),\quad \forall\,r;$$
  \item if $\{ n_{1,0}, n_{1,1}, \ldots, n_{1,N_1}\} \supseteq\{
    n_{2,0}, n_{2,1}, \ldots, n_{2,N_2} \}$, then
    $$d_{(n_{1,0}, \ldots, n_{1,N_1})}(r) \leq d_{(n_{2,0}, \ldots,
      n_{2,N_2})}(r),\quad \forall\,r.$$
  \end{enumerate}
\end{corollary}

\begin{corollary}\label{coro:dmt-reduce}
  Let us define 
  \begin{equation}\label{eq:pk}
    p_k \defeq
    \begin{cases}
      \tilde{n}_0 & k=0, \\
      \sum_{l=0}^k \tilde{n}_l - k \tilde{n}_{k+1} & k=1,\ldots,N-1,\\
      -\infty & k=N.
    \end{cases}
  \end{equation}%
  Then, 
  \begin{equation*}
      d_{({n}_0,\ldots,{n}_N)}(r) = d_{(\tilde{n}_0,\ldots,\tilde{n}_k)}(r),\quad \text{for}\ 
  r\geq p_k.
  \end{equation*}
\end{corollary}
While corollary~\ref{coro:increase} implies that $ d(r) \leq
d_{(\tilde{n}_0,\ldots,\tilde{n}_k)}(r)$ in a general way,
corollary~\ref{coro:dmt-reduce} states precisely that $d(r)$ coincides
with $d_{(\tilde{n}_0,\ldots,\tilde{n}_k)}(r)$ for $r\geq p_k$.

\begin{figure}[!t]
\begin{center}
  \includegraphics[angle=0,width=0.6\textwidth]{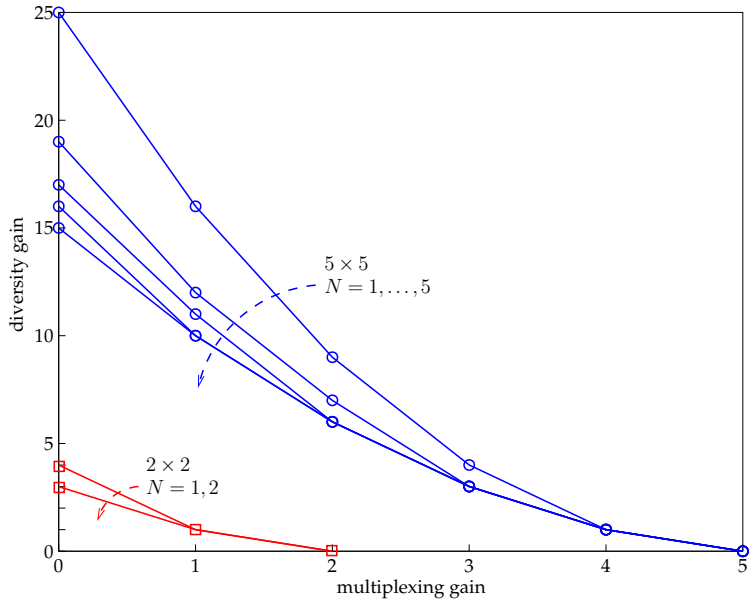}
\caption{Diversity-multiplexing tradeoff of $2\times2$ and $5\times5$ symmetric Rayleigh product channels.}
\label{fig:RayleighProduct_n5}  
\end{center}
\end{figure}

\newcommand{\dUB}{d_{\text{UB}}} \newcommand{\dLB}{d_{\text{LB}}}
\begin{corollary}[Upper bound and lower bound]\label{coro:ub-lb}
\begin{equation*}
  \frac{\tilde{n}_0 \tilde{n}_1}{2} 
  < 
  d(0)
  \leq 
  \tilde{n}_0 \tilde{n}_1
\end{equation*}%
where $d(0)$ is known as the maximum diversity gain.
\end{corollary}
From \Eq{eq:ci} and \Eq{eq:dk}, the upper bound is obtained by setting
$\tilde{n}_2$ large enough and the lower bound is obtain by setting
$\tilde{n}_2=\ldots=\tilde{n}_N$. This corollary implies that the
diversity of a Rayleigh product channel can always be written as
$d(0)=a\tilde{n}_0 \tilde{n}_1$ with $a\in(0.5,1]$. Hence, the
diversity ``bottleneck'' of the Rayleigh product channel $\RP$ is not
necessarily one of the subchannels $\mH_i$, but rather the virtual
$\tilde{n}_0\times\tilde{n}_1$ Rayleigh channel.  On the other hand,
the maximum diversity gain is always strictly larger than
$\frac{\tilde{n}_0 \tilde{n}_1}{2}$, independent of the value $N$.  In
order to illuminate the impact of $N$ on the DMT, let us consider the
symmetric case.
\begin{corollary}[Symmetric Rayleigh product channels]
  When $n_0=n_1=\ldots=n_N=n$, we have
  \begin{equation}\label{eq:dmt_sym}
    d(k) = \frac{(n-k)(n+1-k)}{2} + \frac{a(k)}{2}((a(k)-1)N+2b(k))
  \end{equation}%
  where $a(k) \defeq \left\lfloor \frac{n-k}{N} \right\rfloor$ and
  $b(k) \defeq (n-k)\ \text{mod}\ N$. 
\end{corollary}
In the symmetric case, on one hand, we observe that the DMT degrades
with $N$. On the other hand, from \Eq{eq:dmt_sym}, the degradation
stops at $N=n$ and we have
\begin{equation*}
  d(k) = \frac{(n-k)(n+1-k)}{2}
\end{equation*}%
for $N\geq n$. This can also be deduced from
theorem~\ref{thm:reduction} applying which we get that the order of all symmetric
Rayleigh product channel with $N > n$ is $N^*=n$.
Therefore, we lose less than half of the diversity gain due to the
product of Rayleigh MIMO channels, in contrast to the intuition that
the maximum diversity gain could degrade to $1$ with $N\to\infty$. As
an example, in \Fig{fig:RayleighProduct_n5}, we show the DMT of the
$2\times2$ and $5\times 5$ Rayleigh product channels with different
values of $N$.

\subsection{General Rayleigh Product Channel}
In fact, we can define a more general Rayleigh product channel as
\begin{equation}
  \label{eq:general-RP}
  \RP_g \defeq \mH_1 \mT_{1,2} \mH_2\cdots \mH_{N-1} \mT_{N-1,N}
\mH_N.  
\end{equation}

\begin{theorem}\label{thm:pdf-general-RP}
  The general Rayleigh product channel is equivalent to
  \begin{enumerate}
  \item a $(n_0,n_1,\ldots,n_N)$ Rayleigh product channel, if all the
    matrices $\mT_{i,i+1}$'s are square and their singular values
    satisfy $\sigma_j(\mT_{i,i+1})\asympteq \SNR^0$, $\forall i,j$;
  \item a $(n_0,n'_1,\ldots,n'_{N-1}, n_N)$ Rayleigh product channel,
    with $n'_i$ being the rank of the matrix $\mT_{i,i+1}$, if the
    matrices $\mT_{i,i+1}$'s are constant.
  \end{enumerate}
\end{theorem}
Therefore, the results obtained previously for the Rayleigh product
channel can be applied to the general one.

\section{Amplify-and-Forward Multihop Channels}
\label{sec:af}

Using the results from the previous section, we are going to analyze
the performance of the AF scheme presented in
section~\ref{sec:system-model}, in terms of the DMT.

\subsection{Equivalence to the Rayleigh Product Channel}

With the AF scheme, the end-to-end equivalent MIMO channel is
\begin{equation}\label{eq:multihop_AF1}
  \my_N = \left(\prod_{i=1}^N \mD_i \mH_i\right) \mx_1 + \sum_{j=1}^N \left( \prod_{i=j}^N \mH_{i+1}\mD_i \right) \mz_j
\end{equation}%
where for the sake of simplicity, we define $\prod_{i=1}^N \mA_i
\defeq \mA_N \cdots \mA_1$ for any matrices $\mA_i$'s;
$\mH_{N+1}\defeq\Id$ and $\mD_N\defeq \Id$.  The standard whitened
form of this channel is
\begin{equation*}
  \my = \sqrt{\mR} \left(\prod_{i=1}^N \mD_i \mH_i\right) \mx_1 + \mz
\end{equation*}%
where $\mz\sim\CN$ is the whitened version of the noise and
$\sqrt{\mR}$ is the whitening matrix with $\mR$ the covariance matrix
of the noise in \Eq{eq:multihop_AF1}. Since it can be shown that
$\lambda_{\max}(\mR)\asympteq \lambda_{\min}(\mR) \asympteq \SNR^0$,
the AF multihop channel is DMT-equivalent 
to the channel defined by
$$\mH_N\mD_{N-1}\cdots\mH_2\mD_1\mH_1,$$
which is a general Rayleigh
product channel defined in \Eq{eq:general-RP} if we have
$\sigma_j(\mD_i)\asympteq \SNR^0$, $\forall i,j$. To this end, we
slightly modify the matrices $\mD_i$'s and get the new matrices
$\hat{\mD}_i$ with
$$\hat{\mD}_i[j,j] = \min\left\{\mD_i[j,j], \kappa \right\}$$
where
$0<\kappa<\infty$ is a constant\footnote{The $\kappa$ is only for
  theoretical proof and is not used in practice, since we can always
  set $\kappa$ a very large constant but independent of $\SNR$. In
  this case, $\hat{\mD}_i=\mD_i$ with probability close to $1$ for
  practical $\SNR$.} independent of $\SNR$. Furthermore, it is obvious
that the power constraint is still satisfied by replacing $\mD_i$ with
$\hat{\mD}_i$. Therefore, the multihop channel with the thus defined
AF strategy is DMT-equivalent to a $(n_0,n_1,\ldots,n_N)$ Rayleigh
product channel, \ie,
$$d^{\text{AF}}(k) = \sum_{i=k+1}^{\nmin} c_i.$$
In the rest of the
paper, we identify the Rayleigh product channel, the AF multihop
channel and the vector $(n_0,n_1,\ldots,n_N)$ when confusion is not
likely.

\newcommand{\umH}{\underline{\mH}}
\newcommand{\umD}{\underline{\mD}}
\newcommand{\umQ}{\underline{\mQ}}
\newcommand{\umG}{\underline{\mG}}
\newcommand{\umS}{\underline{\mS}}
\newcommand{\RPPF}{\RP_{\text{PF}}}
\newcommand{\RPPFtran}{\transc{\RP}_{\text{PF}}}
\newcommand{\RPpPF}{\RP'_{\text{PF}}}
\newcommand{\RPpPFtran}{\transc{\left(\RP'_{\text{PF}}\right)}}

\subsection{A Variant~: Project-and-Forward}

\begin{figure*}[!t]
  \centering
  \includegraphics[angle=0,height=0.15\textwidth]{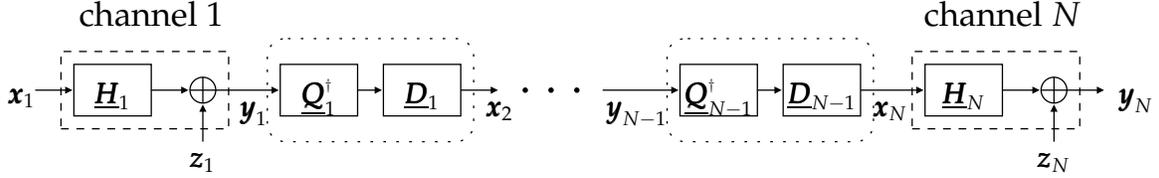}
  \caption{The project-and-forward scheme.}
  \label{fig:multihop_PF}  
\end{figure*}

We propose a new scheme called project-and-forward~(PF), as shown in
\Fig{fig:multihop_PF}. This scheme can be used only when full antenna
cooperation within cluster is possible, that is, all antennas in the
same cluster are controlled by a central unit. At the \node{i}, the
received signal is first projected to the signal subspace $\Scal_i$,
spanned by the columns of the channel matrix $\umH_i$. The dimension
of $\Scal_i$ is $r_i$, the
rank of $\umH_i$.  After the component-wise normalization, the projected signal is
transmitted using $r_i$~(out of $n_i$) antennas. It is now clear that
$\umH_{i+1}\in\CC^{n_{i+1}\times r_i}$ is actually composed of the
$r_i$ columns of the previously defined $\mH_{i+1}$, with $r_0\defeq
n_0$. 

More precisely, the $\umQ_i\in\CC^{n_i\times r_i}$ is an orthogonal
basis of $\Scal_i$ with $\transc{\umQ}_i \umQ_i=\Id$. We can rewrite
$$\umH_i = \umQ_i \umG_i$$
with $\umG_i\in\CC^{r_i\times r_{i-1}}$.
For simplicity, we let $\umQ_i$ be obtained by the QR decomposition of
$\umH_i$ if $n_i>r_{i-1}$ and be identity matrix if $n_i\leq r_i$.
The main idea of the PF scheme is not to use more antennas than
necessary to forward the signal. Since the useful signal lies only in
the $r_i$-dimensional signal subspace, the projection of the received
signal provides sufficient statistics and reduces the noise power by a
factor $\frac{n_i}{r_i}$. In this case, only $r_i$ antennas are needed
to forward the projected signal. Let us define $\mP_i \defeq
\umD_i\transc{\umQ}_i$.
Then, as in the AF case, the PF multihop channel is DMT-equivalent to
the channel defined by
$$\RPPF = \umH_N\mP_{N-1}\cdots\umH_2\mP_1\umH_1.$$
The following
theorem states that using only $r_i$ out of $n_i$ antennas to forward
the projected signal does not incur any loss of diversity, as compared to the AF scheme.  \begin{theorem}\label{thm:PF}
  The PF multihop channel is DMT-equivalent to a
  $(n_0,n_1,\ldots,n_N)$ Rayleigh product channel.
\end{theorem}
While the PF and AF have the same diversity gain, the PF outperforms
the AF in power gain for two reasons. One reason is, as stated before,
that the projection reduces the average noise power. 
The other reason is that the accumulated noise in the AF case is more
substantial than that in the PF case. This is because in the PF case,
less relay antennas are used than in the AF case. Since the power of
independent noises from different transmit antennas add up at the
receiver side, the accumulated noise in the AF case ``enjoys'' a
larger ``transmit diversity order'' than in the PF case.
We call it the \emph{noise hardening} effect. Some examples will be
given in the section of numerical results.

\subsection{Practical Issues}

\subsubsection{Space-Time Coding}

From the input-output point of view, the multihop channel with AF/PF
protocol is merely a linear MIMO fading channel, for which the
DMT-achieving space-time codes exist. For example, in
\cite{Zheng_Tse}, a Gaussian code is shown to achieve the DMT of a
$n_0\times n_1$ Rayleigh channel if the code length $l\geq n_0+n_1-1$.
This result can easily be extended to a general linear fading channel
and one can show that Gaussian coding is DMT-achieving for any fading
statistics if $l$ is large enough.

Another family of code construction is based on cyclic division
algebra~(CDA). These codes have minimum length~$n_0$ and are commonly
known as the Perfect codes~\cite{Oggier_perfect,Elia_perfect}. They
are DMT-achieving thanks to the so-called non-vanishing
determinant~(NVD) property. It has been shown that they are
approximately universal~\cite{Tavildar,Elia_perfect} since they are
DMT-achieving for all fading statistics. Therefore, we propose to use
the rate-$\tilde{n}_0$ $n_0\times n_0$ Perfect codes. In this case,
the only information that the source need to know is
$\tilde{n}_0$. 

\subsubsection{Antenna Reduction}
\label{sec:antenna_reduction}

In the AF case, provided the number of total available antennas
$(n_0,n_1,\ldots,n_N)$, the vertical reduction result gives an exact
number of necessary antennas at each node in the DMT sense. This
result can be used to reduce the number of transmit and relay
antennas\footnote{Reducing the number of receive antennas does not do
  any good, since more receive antennas always provide larger power
  gain without increasing the complexity.}. If Perfect space-time
codes are used, reducing the number of transmit antennas $n_0$ means
reducing the coding length, \ie, coding delay and decoding complexity,
since the code length is equal to the number of transmit antennas.
For instance, only two transmit antennas are needed in a $(4,2,2,2)$
channel. Therefore, instead of using a $4\times4$ Perfect code the
code length of which is $4$, one can use the Golden
code~\cite{Belfiore_Golden} of length $2$ and still achieve the DMT. 

In fact, less relay antennas also means less relay signaling~(relay
probing, synchronization, etc.) overhead especially when different
antennas are from different relaying terminals~(single-antenna
relays).  Furthermore, using more relay antennas hardens the relayed
noise. This is the same phenomenon as we stated in the PF case.
Therefore, the number of relay antennas at each node should be
restricted to $\bar{n}$ (defined in \Eq{eq:minimum_number}), the
number given by the vertical reduction.

\section{Examples and Numerical Results}
\label{sec:nr}

In this section, we provide some examples of multihop channels and
show the performance of AF scheme with simulation results. In all
cases, we make the same assumptions as in
section~\ref{sec:system-model}.

\subsection{Horizontal and Vertical Reduction}
\begin{figure*}[!t]
\begin{center}
  \includegraphics[angle=0,width=0.9\textwidth]{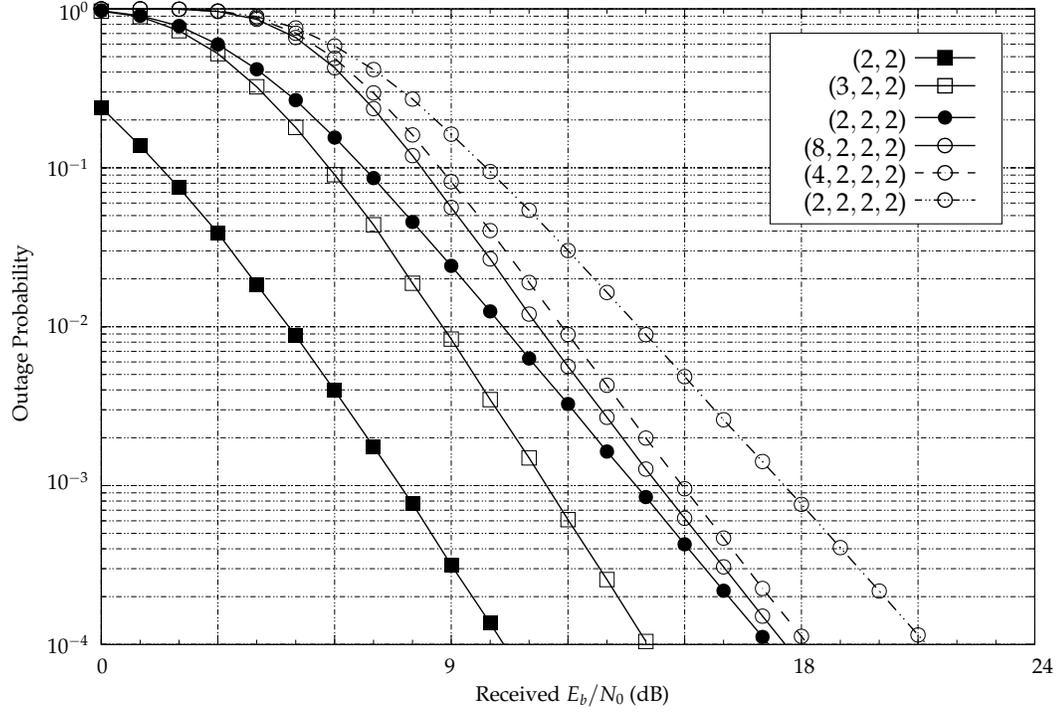}
\caption{Horizontal reduction.}
\label{fig:horizontal_reduction}  
\end{center}
\end{figure*}

Outage performances versus the received SNR per node of different
multihop channels are shown in \Fig{fig:horizontal_reduction}. Note
that both the $(2,2)$ and $(2,2,2)$ channels are minimal and have
diversity order $4$ and $3$, respectively.  The $(3,2,2)$ channel can
be horizontally reduced to $(2,2)$ and thus has diversity $4$.
Similarly, the $(2,2,2,2)$, $(4,2,2,2)$ and $(8,2,2,2)$ channels can
be reduced to $(2,2,2)$ and have diversity $3$. As compared to the
$(2,2,2,2)$ channel, the larger number of transmit antennas in the
$(8,2,2,2)$ weakens the fading of the first hop and the performance is
close to the $(2,2,2)$ channel.

\begin{figure*}[!t]
\begin{center}
  \includegraphics[angle=0,width=0.9\textwidth]{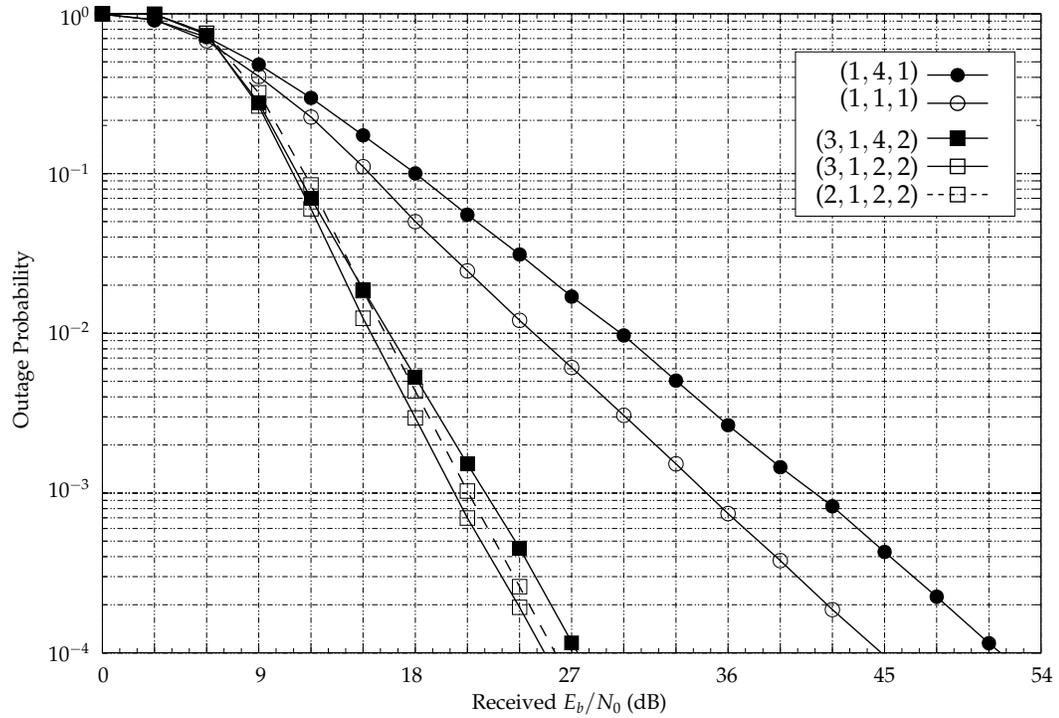}
\caption{Vertical reduction.}
\label{fig:vertical_reduction}  
\end{center}
\end{figure*}

Another example is to illustrate the vertical reduction of multihop
channels, as shown in \Fig{fig:vertical_reduction}. We first consider
the case of a $(1,4,1)$ channel. The necessary antenna number
$\bar{n}$ is $1$ and the minimal vertical form is thus $(1,1,1)$. We
observe that, although both the $(1,4,1)$ and $(1,1,1)$ channels have
diversity $1$, a power gain of $7$~dB is obtained at
$P_{\text{out}}=10^{-4}$ by using only one relay antennas out of four,
if the AF scheme is used. As stated in
section~\ref{sec:antenna_reduction}, the gain is due to avoiding the
hardening of relayed noise. Then, we consider the $(3,1,4,2)$ channel.
The necessary number of antennas $\bar{n}$ is $2$ in this case. As
shown in \Fig{fig:vertical_reduction}, by restricting the number of
relay antennas to $2$, we have a $(3,1,2,2)$ channel and a gain of $2$
dB is observered at $P_{\text{out}}=10^{-4}$. We can further reduce
the number of transmit antennas to $2$ to get a $(2,1,2,2)$ channel.
Unlike the reduction of relay antennas, the reduction of transmit
antennas does not provide any gain because it does not affect the
relayed noise. In contrast, it degrades the performance since the
first hop $(2,1)$ is faded more seriously than the original first hop
$(3,1)$. Nevertheless, the $(2,1,2,2)$ channel is still better than
the $(3,1,4,2)$ channel and is only $0.7$ dB from the $(3,1,2,2)$
channel.

\subsection{Project-and-Forward}

\begin{figure*}[!t]
\begin{center}
  \includegraphics[angle=0,width=0.9\textwidth]{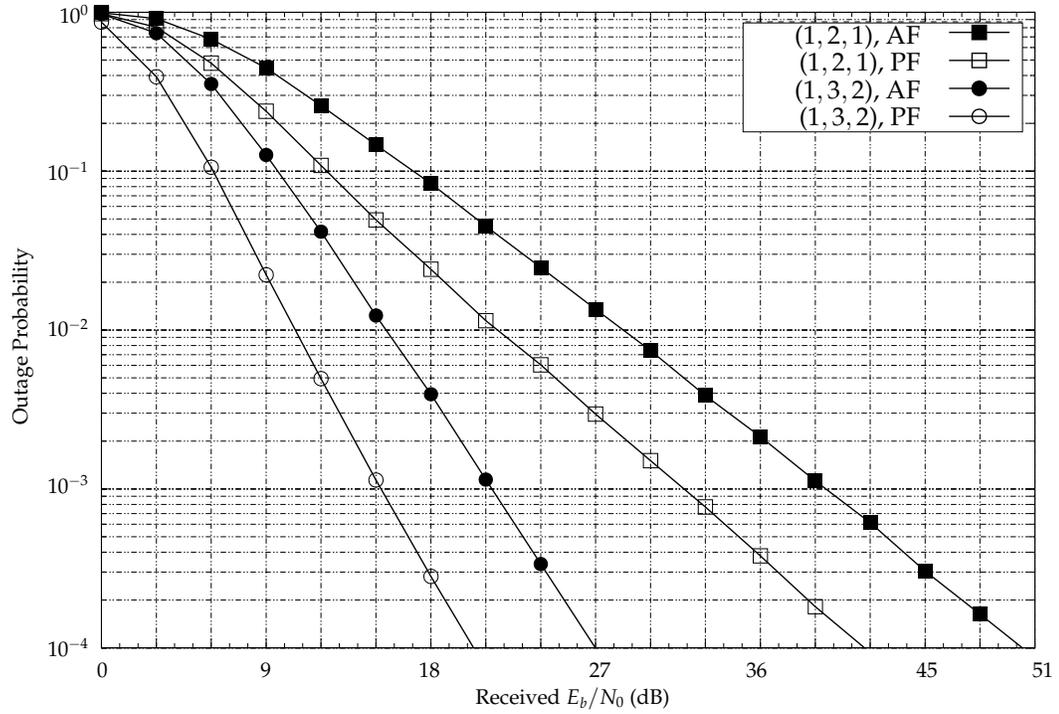}
\caption{AF vs. PF.}
\label{fig:AFvsPF}  
\end{center}
\end{figure*}

In \Fig{fig:AFvsPF}, we compare the PF scheme with the AF scheme for
the $(1,2,1)$ and $(1,3,2)$, respectively. First of all, note that the
AF and the PF have the same diversity order, as predicted. Then, a
power gain of $8.5$~dB~(respectively, $6.5$~dB) over the AF scheme is
obtained by the PF scheme in the $(1,2,1)$~(respective, $(1,3,2)$
channel). This is due to the maximum ratio combining~(MRC) gain in the
first hop and to avoiding the relayed noise hardening.

\subsection{Coded Performance} 

\begin{figure*}[!t]
\begin{center}
  \includegraphics[angle=0,width=0.9\textwidth]{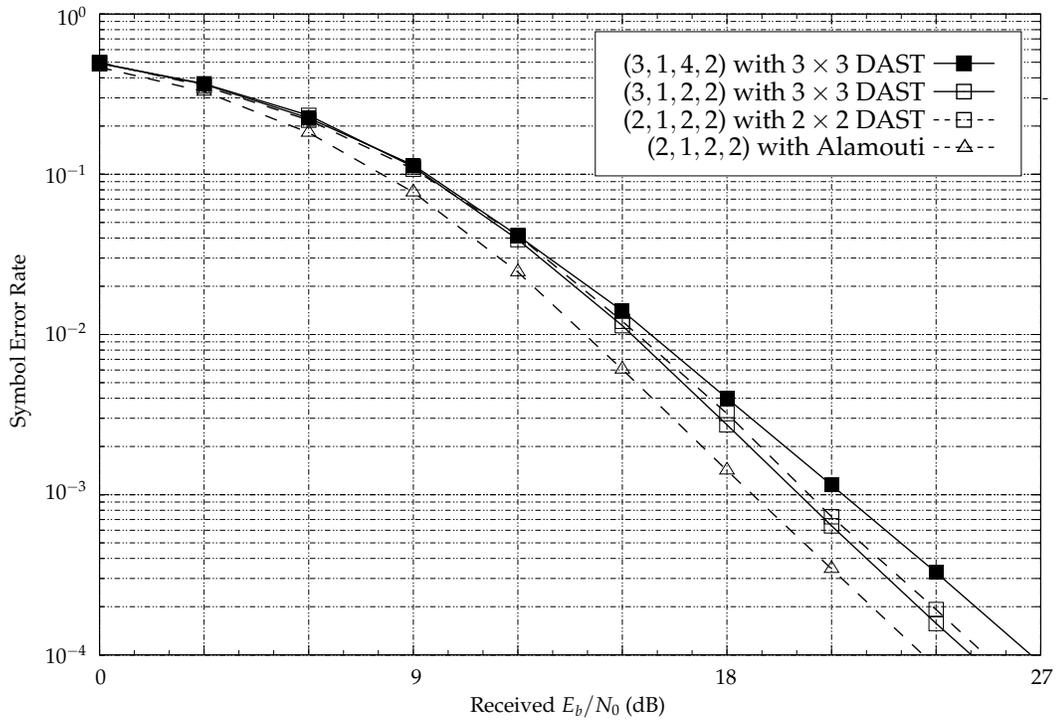}
\caption{Symbol error rate of coded performance.}
\label{fig:coded_performance}  
\end{center}
\end{figure*}

We now study the coded performance of the AF multihop channel. The
performance measure is the symbol error rate~(SER) versus the received
SNR under the maximum likelihood~(ML) decoding. We still take the
$(3,1,4,2)$ channel as an example. Since $\tilde{n}_0 = 1$, the
diagonal algebraic space-time~(DAST) code\footnote{Note that the DAST
  code is the diagonal version of the rate-one Perfect code proposed
  in \cite{Oggier_perfect}.}~\cite{Damen_DAST} can be used. As shown
in \Fig{fig:coded_performance}, with the DAST code, the symbol error
rate performances of in the $(3,1,4,2)$, $(3,1,2,2)$ and $(2,1,2,2)$
channels have exactly the same behavior as the outage performances of
the channels do~\Fig{fig:vertical_reduction}. Moreover, we can use the
Alamouti code~\cite{Alamouti} for the $(2,1,2,2)$ channel. As we can
see in the figure, the Alamouti code outperforms all the DAST codes
with minimum delay and minimum decoding complexity. The potential
benefits from the vertical reduction are thus highlighted.

\subsection{Multihop vs. Direct Transmission}

\begin{figure*}[!t]
\begin{center}
  \subfigure[Total transmission power]{\label{fig:Nhop_2antenna_total}\includegraphics[angle=0,width=0.75\textwidth]{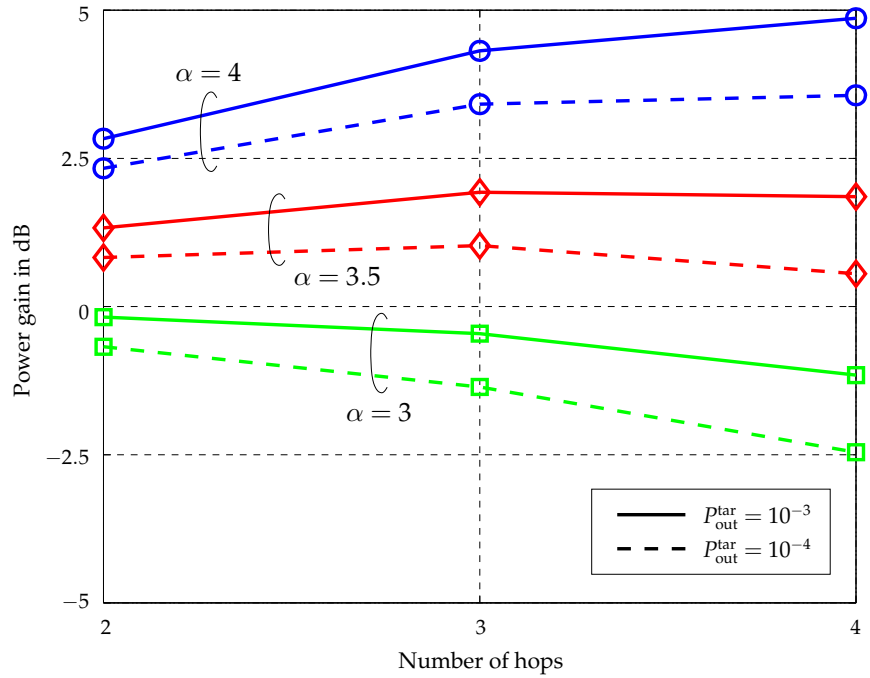}}
\subfigure[Individual transmission power]{\includegraphics[angle=0,width=0.75\textwidth]{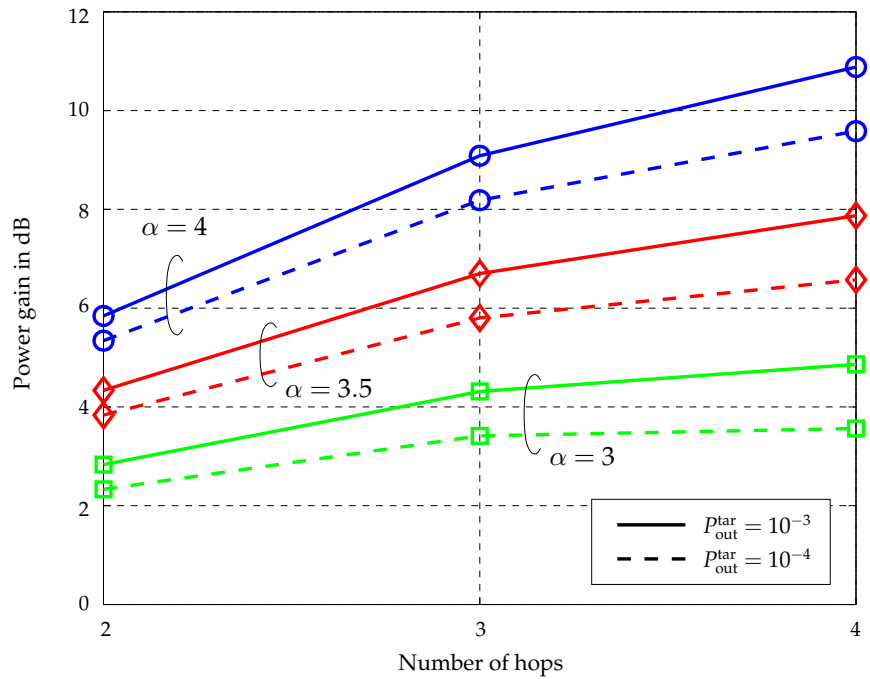}\label{fig:Nhop_2antenna_unit}}
\caption{Transmission power gain of the AF multihop channel over the direct transmission.}
\label{fig:Nhop_2antennas}  
\end{center}
\end{figure*}

Finally, we introduce the path loss model~\cite{Jakes} 
$$
\SNR_{\text{received}} \propto
\text{distance}^{-\alpha}\SNR_{\text{transmitted}} $$
where $\alpha$
is the path loss factor. We fix the distance from the source to the
destination and dispose the relay nodes on the source-destination line
with equal distance. Each node contains two antennas. We compare the
$2$-, $3$- and $4$-hop channel with the direct
transmission~(single-hop) channel. the performance measure is the
transmitted power gain of the multihop channel over the single-hop
channel at certain target outage probability~($10^{-3}$ and
$10^{-4}$). The path loss factor $\alpha$ takes the typical
values~\cite{Jakes} $3$, $3.5$, and $4$ for wireless channels. In
\Fig{fig:Nhop_2antenna_total}, the total transmission power in the
multihop channel is considered. Power gain is obtained for
$\alpha=3.5$ and $4$. Then, the transmission power per node is
considered in \Fig{fig:Nhop_2antenna_unit}. In this case, power gain
is obtained for all $\alpha$ and is as high as $11$~dB. In practice,
the transmission power per node also represents the interference level
for other terminals which has a significant impact on the network
capacity. In both figures, the power gain is lower at $10^{-4}$ than
at $10^{-3}$. This is due to the fact that the direct transmission
channel is a $2\times2$ Rayleigh channel and has diversity $4$, while
the multihop channel is $(2,2,\ldots,2)$ and has diversity $3$. And
low diversity gain means decreasing power gain with increasing SNR or
equivalently, with decreasing outage probability.

\section{Conclusion}
\label{sec:conclusion} 
Perhaps the simplest relaying scheme in the MIMO multihop channel is
the Amplify-and-Forward scheme. In part I of this paper, by
identifying the AF multihop channel with the so-called Rayleigh
product channel, we have obtained the complete characterization of the
diversity-multiplexing tradeoff of the AF scheme in a multihop channel
with arbitrary number of antennas and hops. The characterization is
provided both in direct closed-form and recursive form. Based on the
DMT, a number of properties of the AF multihop channel have been
derived.

In the second part, we will show that the AF scheme is suboptimal in
general, by establishing the diversity upper bound of the multihop
channel with any relaying scheme. By partitioning the multihop channel
into AF subchannels, we achieve the upper bound with both distributed
and non-distributed schemes.

\appendices
\section{Preliminaries}

The followings are some preliminary results that are essential to the
proofs.
\begin{definition}[Wishart Matrix]
  The $m\times m$ random matrix $\mW = \mH\transc{\mH}$ is a (central)
  complex Wishart matrix with $n$ degrees of freedom and covariance
  matrix $\mR$~(denoted as $\mW\sim\Wcal_m(n,\mR)$), if the
  columns of the $m\times n$ matrix $\mH$ are zero-mean independent
  complex Gaussian vectors with covariance matrix $\mR$.
\end{definition}

\begin{lemma}\label{lemma:eq-wishart}
  The joint pdf of the eigenvalues of $\mW \defeq \mH\transc{\mH}
  \sim\Wcal_m(n,\mR_{m\times m})$ is identical to that of any
  $\mW' \sim\Wcal_{m'}(n,\diag(\mu_1,\ldots,\mu_{m'}))$ if
  $\mu_1\geq\ldots\geq\mu_{m'}>\mu_{m'+1}=\ldots=\mu_m = 0$ are the
  eigenvalues of $\mR_{m\times m}$.
\end{lemma}
\begin{proof}
Let $\mR = \transc{\mQ} \diag(\mu_1,\ldots,\mu_{m'},0,\ldots,0) \mQ$
be the eigenvalue decomposition of $\mR$. Then, define
$\sqrt{\mR}\defeq \transc{\mQ}
\diag(\sqrt{\mu_1},\ldots,\sqrt{\mu_{m'}},0,\ldots,0) \mQ$ and
$\mH$ can
be rewritten as $ \mH = \sqrt{\mR} \mH_0 $ with $\mH_0$ having i.i.d.
$\CN[1]$ entries. We know that the eigenvalues of $\mH\transc{\mH}$ are identical to those of 
\begin{align*}
  \transc{\mH}\mH&=\transc{\mH}_0\mR\mH_0 \\
  &=\transc{(\mQ\mH_0)} \diag(\mu_1,\ldots,\mu_{m'},0,\ldots,0) (\mQ
  \mH_0) \\
  &=\transc{\mbs{\widetilde{H}}}_0 \diag(\mu_1,\ldots,\mu_{m'},0,\ldots,0) \mbs{\widetilde{H}}_0\\
  &=\transc{\mbs{\widehat{H}}}_0 \diag(\mu_1,\ldots,\mu_{m'})
  \mbs{\widehat{H}}_0
\end{align*}%
where $\mbs{\widetilde{H}}_0 \defeq \mQ \mH_0\in\CC^{m\times n}$ has
i.i.d. entries as $\mH_0$ does; $\mbs{\widehat{H}}_0\in\CC^{m'\times
  n}$ is composed of the first $m'$ rows of $\mbs{\widetilde{H}}_0$
and its entries is thus i.i.d. as well. Finally, we prove the lemma
using the fact that the eigenvalues of $\transc{\mbs{\widehat{H}}_0}
\diag(\mu_1,\ldots,\mu_{m'}) \mbs{\widehat{H}}_0$ are identical to
those of 
$$\mW' \defeq (\diag(\sqrt{\mu_1},\ldots,\sqrt{\mu_{m'}}) \mbs{\widehat{H}}_0)
\transc{(\diag(\sqrt{\mu_1},\ldots,\sqrt{\mu_{m'}}) \mbs{\widehat{H}}_0)}.$$
\end{proof}

\begin{lemma}[\!\!\cite{James,Gao_Smith,Simon,Tulino_Verdu}]\label{lemma:Wishart}
  Let $\mW$ be a central complex Wishart matrix
  $\mW\sim\Wcal_m(n,\mR)$, where the eigenvalues of $\mR$ are
  distinct\footnote{In the particular case where some eigenvalues of
    $\mR$ are identical, we apply the l'Hospital rule to the pdf
    obtained, as shown in \cite{Simon}.} and their ordered values are
  $\mu_1>\ldots>\mu_m>0$. Let $\lambda_1>\ldots>\lambda_q>0$ be the
  ordered positive eigenvalues of $\mW$ with $q\defeq\min\{m,n\}$. The
  joint pdf of $\mlambda$ conditionned on $\mmu$ is
\begin{subnumcases}{p(\mlambda|\mmu)=} 
  K_{m,n} {\Det(\mXi_1)} \prod_{i=1}^m
  \mu_i^{m-n-1} \lambda_i^{n-m} \prod_{i<j}^m
  \frac{\lambda_i-\lambda_j}{\mu_i-\mu_j}, & \text{if $n\geq m$,}
  \label{eq:Wishart:n>m}
  \\
  G_{m,n} {\Det(\mXi_2)} \prod_{i<j}^m \frac{1}{(\mu_i-\mu_j)}
  \prod_{i<j}^n (\lambda_i-\lambda_j), &\text{if $n<m$,}
  \label{eq:Wishart:n<m}
\end{subnumcases}
with $\mXi_1 \defeq \left[e^{-\lambda_j/\mu_i}\right]_{i,j=1}^m$ and 
  \begin{equation}
    \label{eq:def-Xi2}
    \mXi_2 \defeq \matrix{1 & \mu_1 & \cdots & \mu_1^{m-n-1} & \mu_1^{m-n-1}e^{-\frac{\lambda_1}{\mu_1}} & \cdots & \mu_1^{m-n-1}e^{-\frac{\lambda_n}{\mu_1}} \\
      \vdots & \vdots & \ddots &\vdots& \vdots& \ddots & \vdots \\
        1 & \mu_m & \cdots & \mu_m^{m-n-1} & \mu_m^{m-n-1}e^{-\frac{\lambda_1}{\mu_m}} & \cdots & \mu_m^{m-n-1}e^{-\frac{\lambda_n}{\mu_m}}
}.
  \end{equation}%
  $K_{m,n}$ and $G_{m,n}$ are normalization factors. In particular,
  for $\mR=\Id$, the joint pdf is
  \begin{equation}
    \label{eq:Wishart:Id}
    P_{m,n}  e^{-\sum_i \lambda_i} \prod_{i=1}^q \lambda_i^{\Abs{m-n}} \prod_{i<j}^q (\lambda_i-\lambda_j)^2.    
  \end{equation}
\end{lemma}

Now, let us define the eigen-exponents $\alpha_i\defeq-\log
\lambda_i/\log\SNR,\ i=1,\ldots,q,$ and
$\beta_i\defeq-\log\mu_i/\log\SNR,\ i=1,\ldots,m.$

\begin{lemma}\label{lemma:Det}
  \eqncasesasymptlabel{\Det(\mXi_1)}{\SNR^{-\E_{\mXi_1}(\malpha,\mbeta)}}{for
    $(\malpha,\mbeta)\in\Rcal^{(1)}$}{\SNR^{-\infty}}{otherwise,}{eq:detexp}
where 
\begin{equation}
  \label{eq:EmXi1}
  \E_{\mXi_1}(\malpha,\mbeta) \defeq  \sum_{j=1}^m \sum_{i<j} (\alpha_i-\beta_j)^+,
\end{equation}
and
\begin{equation}
  \label{eq:R1}
  \Rcal^{(1)} \defeq \left\{\alpha_1\leq\ldots\leq\alpha_m,\ \beta_1\leq\ldots\leq\beta_m,\ \text{and}\ \beta_i\leq\alpha_i,\ \text{for}\ i=1,\ldots,m \right\}.
\end{equation}%
\end{lemma}

\begin{proof}

Please refer to \cite{SY_JCB_coop} for details.
\end{proof}

\begin{lemma}\label{lemma:DetXi}
  \eqncasesasymptlabel{\Det\left(\mXi_2\right)}{\SNR^{-\E_{\mXi_2}(\malpha,\mbeta)}}{for
    $(\malpha,\mbeta)\in
    \Rcal^{(2)}$}{\SNR^{-\infty}}{otherwise,}{eq:lemma2} where
  \begin{equation}
    \label{eq:EmXi2}
    \E_{\mXi_2}(\malpha,\mbeta) \defeq \sum_{i=1}^n
        (m-n-1)\beta_i + \sum_{i=n+1}^m (m-i)\beta_i +
        \sum_{j=1}^n\sum_{i<j}\pstv{(\alpha_i-\beta_j)} +
        \sum_{j=n+1}^m\sum_{i=1}^n\pstv{(\alpha_i-\beta_j)}
  \end{equation}
and
  \begin{equation}
    \label{eq:R2}
    \Rcal^{(2)} \defeq \left\{\alpha_1\leq\ldots\leq\alpha_n,\ \beta_1\leq\ldots\leq\beta_m,\ \text{and}\ \beta_i\leq\alpha_i,\ \text{for}\ i=1,\ldots,n \right\}.
  \end{equation}%
\end{lemma}

\begin{proof}
First, we have 
\begin{equation}\label{eq:lemmas:tmp1}
  \Det{(\mXi_2)} = \prod_{i=1}^m \mu_i^{m-n-1} 
  \Det \matrix{
    \mu_1^{-(m-n-1)} &\cdots& 1 & e^{-\lambda_1/\mu_1}&\cdots&e^{-\lambda_n/\mu_1}\\
    \vdots & \ddots & \vdots& \vdots& \ddots & \vdots \\
    \mu_m^{-(m-n-1)} &\cdots& 1 & e^{-\lambda_1/\mu_m}&\cdots&e^{-\lambda_n/\mu_m}\\
  }.
\end{equation}
Then, let us denote the determinant in the right hand side~(RHS) of
\Eq{eq:lemmas:tmp1} as $D$ and we rewrite it as
\begin{align}
  D                                 %
  &= \Det \matrix{
    d_{1,m}^{(m-n-1)} &\cdots& 0 & e^{-\lambda_1/\mu_1}-e^{-\lambda_1/\mu_m}&\cdots&e^{-\lambda_n/\mu_1}-e^{-\lambda_n/\mu_m}\\
    \vdots & \ddots & \vdots& \vdots& \ddots & \vdots \\
    d_{m-1,m}^{(m-n-1)} &\cdots& 0 & e^{-\lambda_1/\mu_{m-1}}-e^{-\lambda_1/\mu_{m}}&\cdots&e^{-\lambda_n/\mu_{m-1}}-e^{-\lambda_n/\mu_{m}}\\
    \mu_m^{-(m-n-1)} &\cdots& 1 & e^{-\lambda_1/\mu_m}&\cdots&e^{-\lambda_n/\mu_m}\\
  } \label{eq:lemmas:tmp3} \\
  &\asympteq \Det \matrix{
    d_{1,m}^{(m-n-1)} &\cdots& d_{1,m}^{(1)} & e^{-\lambda_1/\mu_1}&\cdots&e^{-\lambda_n/\mu_1}\\
    \vdots & \ddots & \vdots& \vdots& \ddots & \vdots \\
    d_{m-1,m}^{(m-n-1)} &\cdots& d_{m-1,m}^{(1)} & e^{-\lambda_1/\mu_{m-1}}&\cdots&e^{-\lambda_n/\mu_{m-1}}\\
  } \prod_{i=1}^n \left(1-e^{-\lambda_i/\mu_m}\right)
  \label{eq:lemmas:tmp2}
\end{align}%
where $d_{i,j}^{(k)}\defeq \mu_i^{-k} - \mu_j^{-k}$ and the product
term in \Eq{eq:lemmas:tmp2} is obtained since
$1-e^{-(\lambda_i/\mu_m-\lambda_i/\mu_j)}\asympteq
1-e^{-\lambda_i/\mu_m}$ for all $j<m$. Let us denote the determinant
in \Eq{eq:lemmas:tmp2} as $D_m$. Then, by multiplying the first column
in $D_m$ with $\mu_m^{m-n-1}$ and noting that $\mu_m^{m-n-1}
d_{i,m}^{(m-n-1)}=1-\left({\mu_m}/{\mu_i}\right)^{m-n-1}\approx
1$, the first column of $D_m$ becomes all $1$. Now, by eliminating the
first $m-2$ ``$1$''s of the first column by subtracting all rows by
the last row as in \Eq{eq:lemmas:tmp3} and \Eq{eq:lemmas:tmp2}, we
have $\mu_m^{m-n-1} D_m\asympteq \prod_{i=1}^n
\left(1-e^{-\lambda_i/\mu_m}\right) D_{m-1}$. By continuing reducing
the dimension, we get
\begin{equation*}
  \begin{split}
    \Det(\mXi_2) &\asympteq  \Det\left[e^{-\lambda_j/\mu_i}\right]_{i,j=1}^n \prod_{i=1}^{n+1}\mu_i^{m-n-1}\prod_{i=n+2}^m \mu_i^{m-i}\\
    &\quad \cdot\prod_{i=1}^n\prod_{j=n+1}^m
    \left(1-e^{-\lambda_i/\mu_j}\right)
  \end{split}
\end{equation*}
from which we prove the lemma, by applying \Eq{eq:detexp}.    
\end{proof}

With the two preceding lemmas, we have the following lemma that
provides the asymptotical pdf of $\malpha$ conditionned on $\mbeta$ in
the high SNR regime.
\begin{lemma}\label{lemma:pdfRayleighCond}
  \eqncasesasympt{p(\malpha|\mbeta)}{\SNR^{-\E(\malpha|\mbeta)}}{for
    $(\malpha,\mbeta)\in\Rcal_{\malpha|\mbeta}$,}{\SNR^{-\infty}}{otherwise,}
where 
\begin{equation}
  \label{eq:expcond}
  \E(\malpha|\mbeta) \defeq \sum_{i=1}^q (n+1-i) \alpha_i + \sum_{i=1}^q (i-n-1)\beta_i
                 + \sum_{j=1}^q\sum_{i<j} (\alpha_i-\beta_j)^+ 
                 + \sum_{j=q+1}^m\sum_{i=1}^q (\alpha_i-\beta_j)^+, 
\end{equation}
and
\begin{equation}
  \label{eq:feasibility}
  \Rcal_{\malpha|\mbeta} \defeq \left\{\alpha_1\leq\ldots\leq\alpha_q,\ \beta_1\leq\ldots\leq\beta_m,\ \text{and}\ \beta_i\leq\alpha_i,\ \text{for}\ i=1,\ldots,q \right\}.
\end{equation}%
\end{lemma}

\begin{proof}
For $n\geq m$, applying the variable changes to \Eq{eq:Wishart:n>m},
we have
\begin{equation}
  \label{eq:pdfcond1:n>m}
  \begin{split}
    p(\malpha|\mbeta) &= K_{m,n} (\log\SNR)^{l}\prod_{i=1}^m \SNR^{-(n-m+1)\alpha_i} \SNR^{-(m-n-1)\beta_i} \nnb\\
    &\quad\cdot \prod_{j=1}^m\prod_{i<j} {(\SNR^{-\alpha_i}-\SNR^{-\alpha_j})}{(\SNR^{-\beta_i}-\SNR^{-\beta_j})^{-1}}  \nnb\\
    &\quad \cdot \Det\left[\exp\left(-\SNR^{-(\alpha_j-\beta_i)}\right)\right].
  \end{split}
\end{equation}
The high SNR exponent of the quantity
$\Det\left[\exp\left(-\SNR^{-(\alpha_j-\beta_i)}\right)\right]$ is
calculated in Lemma~\ref{lemma:Det}. From \Eq{eq:detexp}, we only need
to consider $\alpha_i\geq\beta_i,\forall\,i$, so that
$p(\malpha|\mbeta)$ does not decay exponentially. Therefore, we have
\begin{equation}
  \label{eq:pdfcond2:n>m}
  \begin{split}
    p(\malpha|\mbeta) &\asympteq
    \SNR^{-\left(\sum_{i=1}^m(n+1-i)\alpha_i + \sum_{i=1}^m
        (i-n-1)\beta_i + \sum_{j=1}^m\sum_{i<j} (\alpha_i-\beta_j)^+
      \right)},
  \end{split}
\end{equation}%
if $(\malpha,\mbeta)\in \Rcal^{(1)}$ and $p(\malpha|\mbeta)\asympteq
\SNR^{-\infty}$ otherwise.

For $n<m$, with \Eq{eq:Wishart:n<m} and \Eq{eq:lemma2}, we get
\begin{equation}
  \label{eq:pdfcond1:n<m}
  \begin{split}
    p(\malpha|\mbeta) &\asympteq \prod_{i=1}^n \SNR^{-(m-n-1)\beta_i} \prod_{i=n+1}^m \SNR^{-(m-i)\beta_i} \\
    &\quad\cdot \prod_{j=1}^n\prod_{i<j} \SNR^{-(\alpha_i-\beta_j)^+}
    \prod_{j=n+1}^m \prod_{i=1}^n
    \SNR^{-(\alpha_i-\beta_j)^+} 
    \\
    &\quad \cdot \prod_{i=1}^n \SNR^{-(n+1-i)\alpha_i} \prod_{i=1}^m
    \SNR^{(m-i)\beta_i}.
  \end{split}
\end{equation}%
for $(\malpha,\mbeta)\in\Rcal^{(2)}$ and $p(\malpha|\mbeta)\asympteq
\SNR^{-\infty}$ otherwise. Combining the two cases, we prove the
lemma.
\end{proof}

When $\mR=\Id$, \ie, $\mu_1=\ldots=\mu_m=1$, the joint pdf of
$\malpha$ is found in \cite{Zheng_Tse} as shown in the following lemma.
\begin{lemma}\label{lemma:pdfRayleigh}
  \eqncasesasymptlabel{p(\malpha)}{\SNR^{-\sum_{i=1}^q
      (m+n+1-2i)\alpha_i}}{for
    $\malpha\in\Rcal_{\malpha}$,}{\SNR^{-\infty}}{otherwise,}{eq:pdfRayleigh}
with $\Rcal_{\malpha}\defeq\left\{0\leq\alpha_1\leq\ldots\leq\alpha_q\right\}$.
\end{lemma}
This lemma can be justified either by using \Eq{eq:Wishart:Id} or by
setting $\beta_i=0,\ \forall\,i$ in \Eq{eq:expcond}.

\begin{lemma}[\cite{SY_JCB_ds}]\label{lemma:invariance-asymp}
  Let $\mM$ be any $m\times n$ random matrix and $\mT$ be any $m\times
  m$ non-singular matrix whose singular values satisfy
  $\sigma_{\min}(\mT)\asympteq\sigma_{\max}(\mT)\asympteq\SNR^0$.
  Define $q\defeq\min\{m,n\}$ and $\mbs{\tilde{M}} \defeq \mT\mM$. Let
  $\sigma_1(\mM)\geq\ldots\geq\sigma_q(\mM)$ and
  $\sigma_1(\mbs{\tilde{M}})\geq\ldots\geq\sigma_q(\mbs{\tilde{M}})$
  be the ordered singular values of $\mM$ and $\mbs{\tilde{M}}$,
  Then, we have
  \begin{equation*}
    \sigma_i(\mbs{\tilde{M}}) \asympteq \sigma_i(\mM),\quad\forall i.
  \end{equation*}%
\end{lemma}

\section{Proof of Theorem~\ref{thm:asympt-pdf}}\label{sec:proof-thm-pdf}
The following lemma will be used repeatedly in the most of the proofs.
\begin{lemma}~\label{lemma:ci}
  Let $\Ical_k\defeq[\,p_k,p_{k-1}]$, $k=1,\ldots,N$, be $N$
  consecutively joint intervals with $p_N\defeq -\infty$, $p_0 \defeq
  \tilde{n}_0$, and $p_k$'s are defined as in \Eq{eq:pk}. 
  Then, we have
  \begin{equation}\label{eq:ci3}
    c_i = 1-i + \left\lfloor\frac{\sum_{l=0}^{k}\tilde{n}_l - i}{k} \right\rfloor, \quad \text{for}\ i\in\Ical_{k}.
  \end{equation}%
\end{lemma}
\begin{proof}
  $c_i$ defined by \Eq{eq:ci} is the minimum of $N$ sequences
  corresponding to the $N$ values of $k$. It is enough to show that
  each of the $N$ sequences dominates in a consecutive manner. We omit
  the details here.
\end{proof}

\subsection{Sketch of the Proof}
\label{sec:sketch-proof}
The proof will be by induction on $N$. From
lemma~\ref{lemma:pdfRayleigh}, the theorem is trivial for $N=1$. 
Suppose the theorem holds for some $N$ and
$\RP\defeq\mH_1\cdots\mH_N$, we would like to show that it is also
true for $N+1$ and $\RP' \defeq \mH_{1}\cdots\mH_{N+1}$. For
simplicity, the ``primed'' notations~(\eg, $\malpha'$, $\mn'$,
$\mnt'$, $\mc'$, $\nminp$, etc.) will be used for the respective
parameters of $\RP'$.  Note that
$\RP'\transc{(\RP')}\sim\Wcal_{n_0}(n_{N+1},\RP\transc{\RP})$ for a
given $\RP$, since $\RP' = \RP \mH_{N+1}$.  According to
lemma~\ref{lemma:eq-wishart}, the pdf of the eigenvalues $\mlambda'$
of $\RP'\transc{(\RP')}$ is identical to that of
$\Wcal_{\nmin}(n_{N+1},\diag(\mlambda))$.  Hence, the pdf of
$\malpha'$ can be obtained as the marginal pdf of $(\malpha',\malpha)$
\begin{align}
  p(\malpha') &= \int_{\RR^{{\nmin}}} p(\malpha',\malpha) \d \malpha \nonumber\\
  &= \int_{\RR^{{\nmin}}} p(\malpha'|\malpha) p(\malpha) \d \malpha \nonumber\\
  &\asympteq \int_{\Rcal} \SNR^{-\E(\malpha'|\malpha)}
  \SNR^{-\E(\malpha)}
  \d \malpha \label{eq:tmp31}\\
  &\asympteq \SNR^{-\Ec(\malpha')} \label{eq:tmp32}
\end{align}%
where \Eq{eq:tmp31} comes from lemma~\ref{lemma:pdfRayleighCond} and
our assumption that \Eq{eq:p-malpha} holds for $N$, with
\begin{align}
  \Rcal
  &\defeq \Rcal_{\malpha'|\malpha}\cap\Rcal_{\malpha} \nonumber\\
  &=\left\{0\leq\alpha'_1\leq\ldots\leq\alpha'_{\nminp},\ 
    0\leq\alpha_1\leq\ldots\leq\alpha_{\nmin},\ \text{and}\ 
    \alpha_i\leq\alpha'_i,\ \text{for}\ i=1,\ldots,\nminp \right\} \label{eq:feasibleRegion}
\end{align}%
being the feasible region; the exponent $\Ec(\malpha')$ in
\Eq{eq:tmp32} is defined by
\begin{align}
  \Ec(\malpha') &= \min_{\malpha\in\Rcal} \E(\malpha',\malpha)
  \label{eq:minprob}
\end{align}%
with $\E(\malpha',\malpha) \defeq \E(\malpha'|\malpha) +
\E(\malpha)$. From \Eq{eq:expcond} and \Eq{eq:Ea},
\begin{multline}
  \E(\malpha',\malpha) = \sum_{i=1}^{\nminp} (n_{N+1}-i+1) \alpha_i'
  + \sum_{j=1}^{\nminp} \left( (j-1-n_{N+1}+c_j)\alpha_j + \sum_{i<j}\pstv{(\alpha_i'-\alpha_j)} \right) \\
  + \sum_{j=\nminp+1}^{{\nmin}} \left( c_j \alpha_j +
    \sum_{i=1}^{\nminp} \pstv{(\alpha_i'-\alpha_j)} \right).
  \label{eq:jointexp}
\end{multline}%
It remains to show $\Ec(\malpha')=\E'(\malpha')\defeq
\sum_i c_i \alpha'_i$ with
\begin{equation}\label{eq:ci2}
  c'_i \defeq 1-i + \min_{k=1,\ldots,N+1} \left\lfloor\frac{\sum_{l=0}^{k}\tilde{n}'_l - i}{k} \right\rfloor,\quad i=1,\ldots,\nminp
\end{equation}%
by solving the optimization problem \Eq{eq:minprob}, which is
accomplished in the rest of the section.


\subsection{Solving the Optimization Problem}
\begin{figure*}
  \subfigure[Case~1]{
  \epsfig{figure=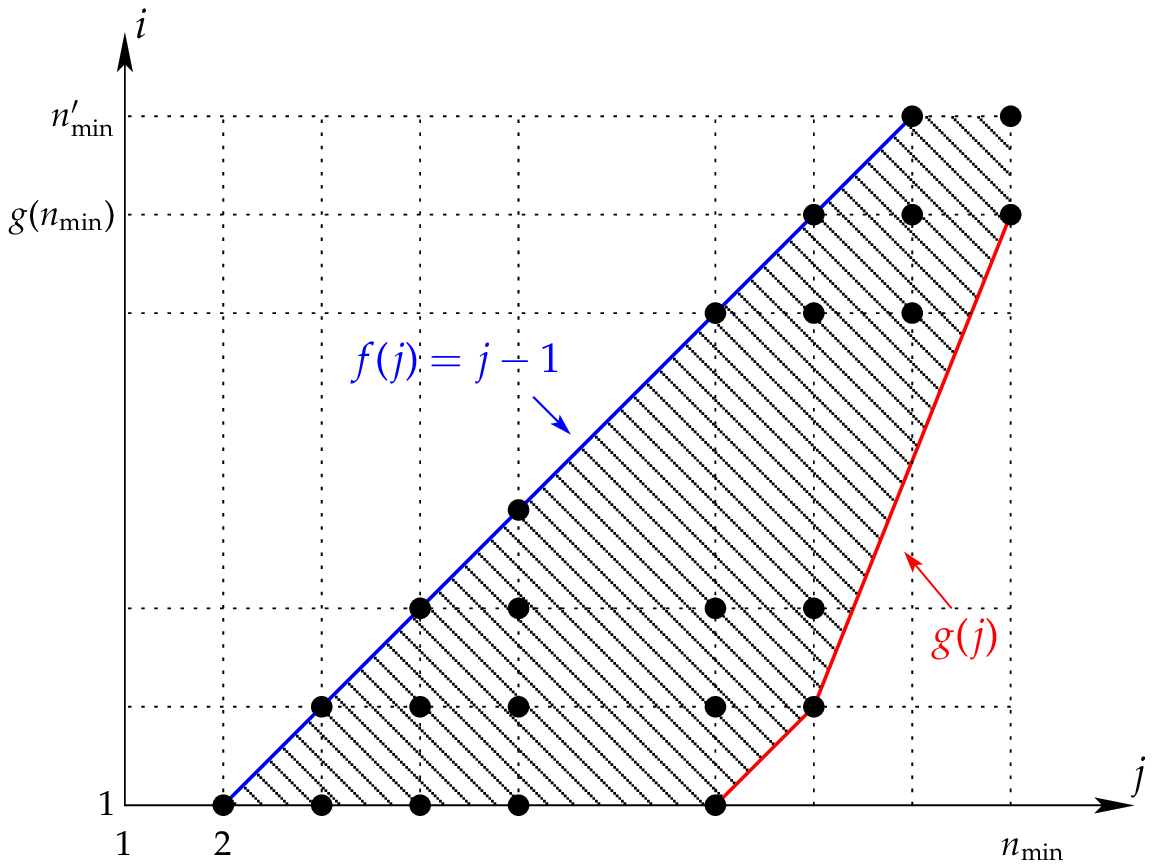,height=0.25\textwidth}
  \label{fig:find-ci1}    }
  \subfigure[Case~2]{
  \epsfig{figure=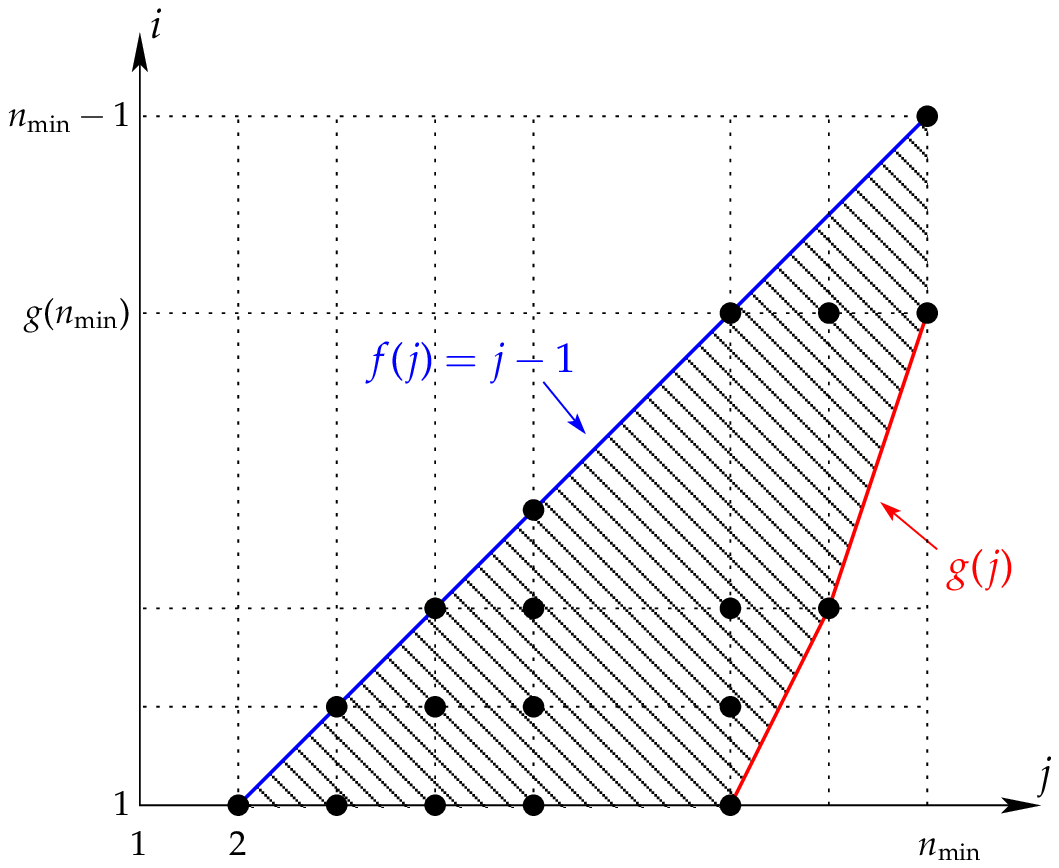,height=0.25\textwidth}
  \label{fig:find-ci2}  }
  \subfigure[Case~3]{
  \epsfig{figure=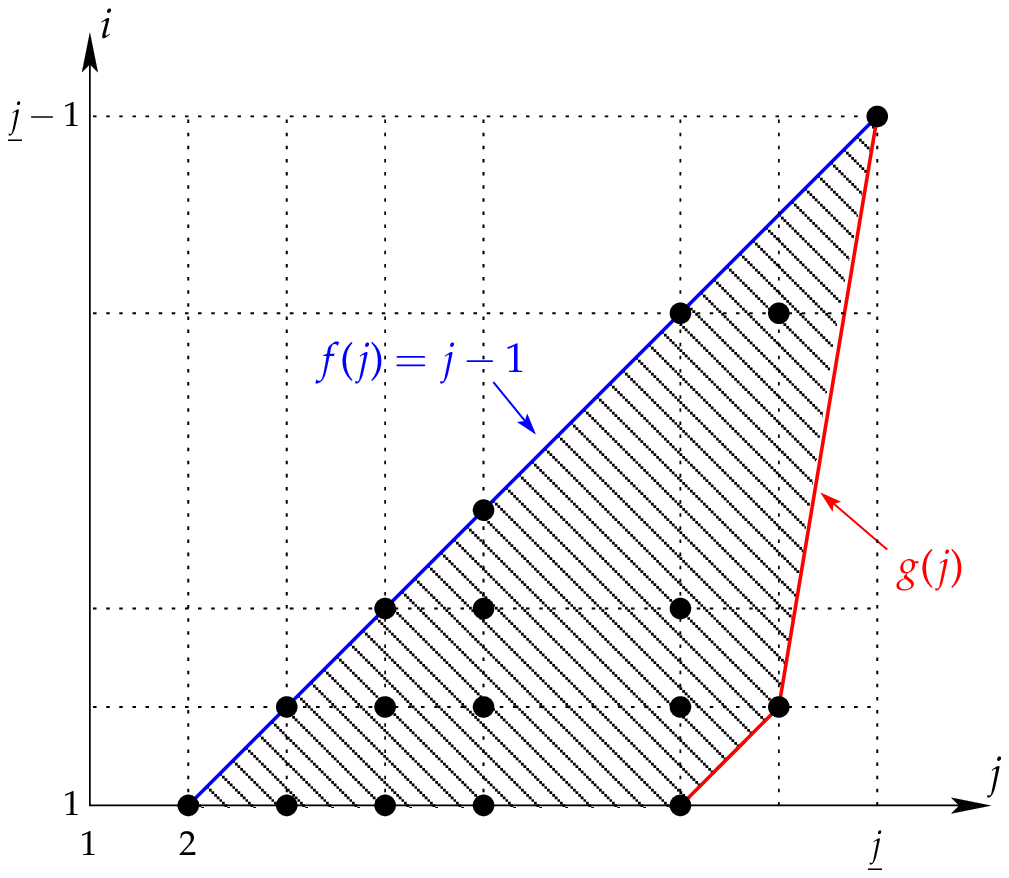,height=0.25\textwidth}
  \label{fig:find-ci3}  }
\caption{For each $j$, the black dots represent the $\alpha'$'s that are freed 
  by $\alpha_j$. Therefore, we can get the total number of freed
  $\alpha'_i$ by counting the black dots in row $i$. More precisely,
  there are $\left\lfloor{g^{-1}(i)}\right\rfloor - \left\lceil{
      f^{-1}(i) }\right\rceil + 1=\left\lfloor{g^{-1}(i)}\right\rfloor
  - i$ black dots for $i\leq g({\nmin})$, and ${\nmin} - \left\lceil{ f^{-1}(i)
    }\right\rceil + 1={\nmin} - i$ black dots for $i > g({\nmin})$.}
\end{figure*}%

\subsubsection{Case 1~[$n_{N+1}<\tilde{n}_0$]}
\label{sec:n_n+1in1-tilden_0}

In this case, we have $\nminp=\tilde{n}'_0=n_{N+1}$. 
Minimization of $\E(\malpha,\malpha')$ of \Eq{eq:jointexp} with
respect to~(\wrt) $\malpha$ can be decomposed into ${\nmin}$
minimizations \wrt~$\alpha_1,\ldots,\alpha_{{\nmin}}$ successively,
\ie,
\begin{equation*}
  \min_{\malpha} = \min_{\alpha_{\nmin}}\cdots\min_{\alpha_1}.
\end{equation*}%
We start with $\alpha_1$. From \Eq{eq:feasibility}, the feasible
region of $\alpha_1$ is $0\leq\alpha_1\leq\alpha'_1$.  Since the only
$\alpha_1$-related term in \Eq{eq:jointexp} is $(c_1-n_{N+1})\alpha_1$
and $c_1-n_{N+1}>0$ for $n_{N+1}<\tilde{n}_0$, we have $\alpha_1^* =
0$.  Now, suppose that the minimization
\wrt~$\alpha_1,\ldots,\alpha_{j-1}$ is done and that we would like to
minimize \wrt~$\alpha_j$. For $\alpha_j$, $j\leq \nminp$, we set the
initial region as
\begin{equation*}
  0\leq\alpha'_1\leq\cdots\leq\alpha'_{j-1}\leq\alpha_j\leq\alpha'_j
\end{equation*}%
in which we have $\sum_{i<j}\pstv{(\alpha'_i-\alpha_j)}=0$. The
feasibility conditions in \Eq{eq:feasibleRegion} require that
$\alpha_j$ must not go right across $\alpha'_j$. The only choice is
therefore to go to the left. Each time $\alpha_j$ goes across an
$\alpha'_i$ from the right to the left, $(\alpha'_i-\alpha_j)^+$
increases by $\alpha'_i-\alpha_j$, which increases the coefficient of
$\alpha'_i$ by $1$ and decreases the coefficient of $\alpha_j$ by $1$.
It can be shown that, to minimize the value of $\E(\malpha,\malpha')$
\wrt~$\alpha_j$, $\alpha_j$ is allowed to cross $\alpha'_i$ only when
the current coefficient of $\alpha_j$ in \Eq{eq:jointexp} is
positive\footnote{When the coefficient of $\alpha_i$ in
  \Eq{eq:jointexp} is positive, decreasing $\alpha_i$ decreases
  $\E(\malpha,\malpha')$.}. So, $\alpha_j$ stops moving only in the
following two cases~: 1) it hits the left extreme, $0$; and 2) its
coefficient achieves $0$ when it is in the interval
$[\alpha'_k,\alpha'_{k+1}]$ for some $k<j$. Either case,
$\alpha_j$-related terms are gone and what remain are the
$\alpha'_i$'s ``freed'' by $\alpha_j$ from $\sum_{i<j}
\pstv{(\alpha'_i-\alpha_j)}$. Same reasoning applies to $\alpha_j$ for
$j> \nminp$, except that the initial region is set to
$0\leq\alpha'_1\leq\cdots\leq\alpha'_{\nminp}\leq\alpha_j$.

Therefore, the optimization problem can be solved by counting the
total number of freed $\alpha'_i$'s. As shown in \Fig{fig:find-ci1},
when $j$ is small, the initial coefficient of $\alpha_j$ is large and
thus $\alpha_j$ can free out $\alpha'_{j-1},\ldots, \alpha'_{1}$. We
have $\alpha_j^*=0$, which corresponds to the first stopping
condition. For large $j$, the initial coefficient of $\alpha_j$ is not
large enough and only $\alpha'_{j-1},\ldots,\alpha'_{g(j)}$ is freed,
which corresponds to the second stopping condition. With the above
reasoning, we can get $g(j)$
\eqncaseslabel{g(j)}{j-1-(j-1-n_{N+1}+c_j)+1}{$j\leq
  \nminp$,}{n_{N+1}-c_j+1}{$j>\nminp$.}{eq:tmp48} From \Eq{eq:tmp48}
and \Eq{eq:ci}, we get
\begin{equation}
  \label{eq:gi}
  g(j) = n_{N+1} - \min_{k=1,\ldots,N} \left\lfloor\frac{\sum_{l=0}^{k}\tilde{n}_l - (k+1)j}{k} \right\rfloor,
\end{equation}
and 
\begin{align}
  \left\lfloor g^{-1}(i)\right\rfloor &= \min_{k=1,\ldots,N}
  \left\lfloor\frac{\sum_{l=0}^{k}\tilde{n}_l - k(n_{N+1}-i)}{k+1}
  \right\rfloor \label{eq:gi_inv1}.
\end{align}%
Now, $\Ec(\malpha')$ can be obtained\footnote{In the above
  minimization procedure, we ignored the feasibility condition
  $\alpha_j\geq\alpha_k,\ \forall\,j>k$. A more careful analysis can
  reveal that it is always satisfied with the described procedure.}
from \Fig{fig:find-ci1}
\begin{align}
  \Ec(\malpha') &= \sum_{i=1}^{\nminp} (n_{N+1}-i+1) \alpha'_i +
  \sum_{i=1}^{g({\nmin})} (\left\lfloor
    g^{-1}(i)\right\rfloor-i)\alpha'_i +
  \sum_{i=g({\nmin})+1}^{\nminp}
  (\nmin-i)\alpha'_i \nonumber\\
  &= \sum_{i=1}^{g({\nmin})} \left( 1-2i+ n_{N+1} + \left\lfloor
      g^{-1}(i)\right\rfloor \right) \alpha'_i +
  \sum_{i=g({\nmin})+1}^{\nminp} \left( 1-2i+
    n_{N+1} + \nmin \right) \alpha'_i \nonumber\\
  &= \sum_{i=1}^{g({\nmin})} \left( 1 - i + \min_{k=2,\ldots,N+1}
    \left\lfloor\frac{\sum_{l=0}^{k}\tilde{n}'_l - i}{k}\right\rfloor
  \right) \alpha'_i +
  \sum_{i=g({\nmin})+1}^{\nminp} \left(1-2i+ n_{N+1} + \nmin \right) \alpha'_i\label{eq:tmp89}\\
  &= \sum_{i=1}^{\nminp} \left( 1-i + \min_{k=1,\ldots,N+1}
    \left\lfloor\frac{\sum_{l=0}^{k}\tilde{n}'_l - i}{k}\right\rfloor \right) \alpha'_i \label{eq:tmp54}\\
  &= \E'(\malpha'),
\end{align}%
where \Eq{eq:tmp89} is from \Eq{eq:gi_inv1} and the fact that
$\tilde{n}'_0=n_{N+1}$, $\tilde{n}'_l=\tilde{n}_{l-1}$, $l=1,\ldots,N+1$; \Eq{eq:tmp54}
can be derived from lemma~\ref{lemma:ci}, since $p'_1 =
n_{N+1}+\tilde{n}_0-\tilde{n}_1 = g({\nmin})$ and therefore the term
$\min_k$ in \Eq{eq:tmp54} is dominated by $k\geq2$ for $i\leq g({\nmin})$
and by $k=1$ for $i>g({\nmin})$, corresponding to the two terms in
\Eq{eq:tmp89}, respectively.

\subsubsection{Case 2~[$n_{N+1}\in[\tilde{n}_0,\tilde{n}_1)$]}
\label{sec:n_n+1in1-tilden_1}
In this case, we have $\nminp={\nmin}$ and $\tilde{n}'_1 = n_{N+1}$. From
\Eq{eq:jointexp},
\begin{multline}
  \E(\malpha',\malpha) = \sum_{i=1}^{\nminp} (n_{N+1}-i+1) \alpha_i' +
  \sum_{j=1}^{\nminp} \left( (j-1-n_{N+1}+c_j)\alpha_j + \sum_{i<j}
  \pstv{(\alpha_i'-\alpha_j)}\right). \label{eq:jointexp2}
\end{multline}
Since $j-1-n_{N+1}+c_j>0$, $\forall\,j\leq {\nminp}$, the minimization of
$\E(\malpha',\malpha)$ \wrt~$\malpha$ is in exactly the same manner as
in the previous case. Therefore, $\Ec(\malpha')$ can be obtained from
\Fig{fig:find-ci2} with $g(j)$ in the same form as \Eq{eq:gi}
\begin{align}
  \Ec(\malpha') &= \sum_{i=1}^{\nminp} (n_{N+1}-i+1) \alpha'_i +
  \sum_{i=1}^{g({\nmin})} (\left\lfloor
    g^{-1}(i)\right\rfloor-i)\alpha'_i +
  \sum_{i=g({\nmin})+1}^{\nminp}
  (\nmin-i)\alpha'_i \nonumber\\
  &= \E'(\malpha').
\end{align}%

\subsubsection{Case 3~[$n_{N+1}\in[\tilde{n}_1,\infty)$]}
\label{sec:n_n+1in1-tilden_3}
As in the last case, we have $\nminp={\nmin}$ and the same
$\E(\malpha',\malpha)$ as defined in \Eq{eq:jointexp2}. Without loss
of generality, we assume that
$n_{N+1}\in[\tilde{n}_{k^*},\tilde{n}_{{k^*}+1})$ for some ${k^*}\in[1,N]$~(we set
$\tilde{n}_{N+1} \defeq \infty$). Then, we have
\begin{equation}
  \label{eq:tmp98}
  \tilde{n}'_l = \tilde{n}_l,\quad \text{for}\ l=1,\ldots,{k^*},
\end{equation}%
and
\begin{equation}
  \label{eq:tmp99}
  p_{k^*} < p'_{k^*}\leq p_{{k^*}-1}=p'_{{k^*}-1}\leq\cdots\leq p_1=p'_1.
\end{equation}%
Unlike the previous case, $j-1-n_{N+1}+c_j$ is not always positive.
Let $\jlb$ be the smallest integer such that the coefficient
$j-1-n_{N+1}+c_j$ of $\alpha_j$ in \Eq{eq:jointexp2} is zero. It is
obvious that for $j\geq\jlb$, $\alpha^*_j=\alpha'_j$. Hence, we have
\begin{align}
  \Ec(\malpha') &= \sum_{i=1}^{\nminp} (n_{N+1}-i+1) \alpha'_i +
  \sum_{i=1}^{\jlb-1} (\left\lfloor g^{-1}(i)\right\rfloor-i)\alpha'_i
  + \sum_{j=\jlb}^{\nminp}
  (j-1-n_{N+1}+c_j)\alpha'_j \nonumber
\end{align}%
where the second term is from \Fig{fig:find-ci3}. Furthermore, we can
show that $\jlb \leq p'_{k^*}$, since $p'_{k^*}-1-n_{N+1}+c_{p'_{k^*}}=0$.
Therfore, we get
\begin{align}
  \Ec(\malpha') &= \sum_{i=1}^{\jlb-1} \left( 1-2i+ n_{N+1} +
    \left\lfloor g^{-1}(i)\right\rfloor \right) \alpha'_i +
  \sum_{i=\jlb}^{p'_{k^*}-1} (n_{N+1}-i+1) \alpha'_i +
  \sum_{i=p'_{k^*}}^{\nminp} c_i \alpha'_i. \label{eq:tmp56}
\end{align}%
Now, we would like to show that the coefficient of $\alpha'_i$ in
\Eq{eq:tmp56} coincides with $c'_i$.  First, for $i\leq\jlb-1$, $i\in
\,\Ical'_{{k^*}+1}\cup\cdots\cup\Ical'_N$ and
lemma~\ref{lemma:ci} implies that
\begin{align*}
  1 - 2i + n_{N+1} + \left\lfloor g^{-1}(i)\right\rfloor &= 1 - i +
  \min_{k=2,\ldots,N+1} \left\lfloor\frac{\sum_{l=0}^{k}\tilde{n}'_l -
      i}{k}\right\rfloor \\
  &= 1 - i + \min_{k=1,\ldots,N+1}
  \left\lfloor\frac{\sum_{l=0}^{k}\tilde{n}'_l -
      i}{k}\right\rfloor \\
  &=c'_i.
\end{align*}
Then, for $i\geq p'_{k^*}$, we have 
\begin{equation*}
  i\,\in\,\left(\Ical'_{{k^*}}\cup\cdots\cup\Ical'_1\right)\cap\left(\Ical_{{k^*}}\cup\cdots\cup\Ical_1\right).
\end{equation*}%
Hence, 
\begin{align}
  c'_i &= 1 - i + \min_{k=1,\ldots,k^*}
  \left\lfloor\frac{\sum_{l=0}^{k}\tilde{n}'_l -
      i}{k}\right\rfloor \nonumber\\
       &= 1 - i + \min_{k=1,\ldots,k^*}
  \left\lfloor\frac{\sum_{l=0}^{k}\tilde{n}_l -
      i}{k}\right\rfloor \label{eq:tmp112}\\
       &= c_i, \nonumber
\end{align}
where \Eq{eq:tmp112} is from \Eq{eq:tmp98} and \Eq{eq:tmp99}. Finally,
for $i\in[\jlb,p'_{k^*})$, let us rewrite $i=p'_{k^*}-\Delta_i$. Since
$i-1-n_{N+1}+c_i=0$, $\forall\,i\in[\jlb,p'_{k^*})$, we have
\begin{align*}
  \left\lfloor \frac{\sum_{l=0}^{k^*} \tilde{n}_l - i - {k^*} n_{N+1}}{{k^*}}
  \right\rfloor &= \left\lfloor \frac{\sum_{l=0}^{k^*} \tilde{n}_l - p'_{k^*}
      + \Delta_i - {k^*} n_{N+1}}{{k^*}} \right\rfloor \nonumber\\
  &=  \left\lfloor \frac{\Delta_i}{{k^*}} \right\rfloor \nonumber\\
  &= 0,
\end{align*}%
from which we have $\Delta_i\in[0,{k^*}-1]$ and
\begin{align*}
  c'_i &= \left\lfloor \frac{\sum_{l=0}^{k^*} \tilde{n}_l + n_{N+1}
      -i}{{k^*}+1}\right\rfloor + 1 - i \nonumber\\
  &= \left\lfloor \frac{\sum_{l=0}^{k^*} \tilde{n}_l + n_{N+1}
      - p'_{k^*} + \Delta_i}{{k^*}+1}\right\rfloor + 1 - i \nonumber\\
  &= 1 + n_{N+1} - i.
\end{align*}%
The proof is complete.

\subsection{Proof of Theorem~\ref{thm:pdf-general-RP}}\label{sec:proof-pdf-general-RP}
  To prove the first case, we use induction on $N$. Suppose that it is
  true for $N$, which means that the joint pdf of $\malpha(\RP_g
  \transc{\RP}_g)$ is the same as that of $\malpha(\RP
  \transc{\RP})$. Furthermore, we know by
  lemma~\ref{lemma:invariance-asymp} that $\malpha(\RP_g\mT_{N,N+1}
  \transc{\mT_{N,N+1}}\transc{\RP}_g) = \malpha(\RP_g
  \transc{\RP}_g)$. Same steps as \Eq{eq:tmp31}\Eq{eq:tmp32} complete
  the proof.
  
  To prove the second statement, we perform a singular value
  decomposition on the matrices $\mT_{i,i+1}$'s and then apply the first
  statement.

\section{Proof of Theorem~\ref{thm:reduction} and Theorem~\ref{thm:minimal}}\label{sec:proof-thm-reduction}
\subsection{Proof of Theorem~\ref{thm:reduction}}
Let $$c_i^{(m)} \defeq 1-i + \min_{k=1,\ldots,m}
\left\lfloor\frac{\sum_{l=0}^{k}\tilde{n}_l - i}{k}
\right\rfloor,\quad i=1,\ldots,{\nmin}.$$
What we should prove is that
$$c_i^{(N)}=c_i^{(k)},\quad\text{for}\ i=1,\ldots,\nmin$$
if and only
if \Eq{eq:reduction-cond} is true. To this end, it is enough to show
that
\begin{equation}
  \label{eq:tmp111}
  c_i^{(N)} = c_i^{(N-1)} \quad\text{for}\ i=1,\ldots,\nmin
\end{equation}%
if and only if $p_{N-1}\leq N-1$, that is,
$(N-1)\left(\tilde{n}_N+1\right)\geq \sum_{l=0}^{N-1} \tilde{n}_l$,
and then apply the result successively to show the theorem.

\subsubsection{The Direct Part}
\label{sec:direct-part}

The direct part is to show that, if $p_{N-1}\leq N-1$, then
\Eq{eq:tmp111} is true. From lemma~\ref{lemma:ci}, we see that
$c_i^{(N)}=c_i^{(N-1)},\ \forall\,i\geq p_{N-1}$. Hence, when
$p_{N-1}\leq 1$, \Eq{eq:tmp111} holds. Now, let us consider the case
$p_{N-1}>1$. We would like to show that $c_i^{(N)}=c_i^{(N-1)}$ for
  $i\in[1,p_{N-1}]$. Let $j\defeq p_{N-1}-i \in[0,p_{N-1}-1]$. Then,
  we rewrite the two quantities
  \begin{align}
    \left\lfloor\frac{\sum_{l=0}^{N}\tilde{n}_l - i}{N} \right\rfloor
    &= \tilde{n}_N+\left\lfloor\frac{j}{N}
    \right\rfloor \label{eq:tmp222}\\
    \left\lfloor\frac{\sum_{l=0}^{N-1}\tilde{n}_l - i}{N-1}
    \right\rfloor &= \tilde{n}_N+\left\lfloor\frac{j}{N-1} \right\rfloor \label{eq:tmp333}
  \end{align}%
  that are identical for $p_{N-1}\leq N-1$, which proves that
  $c_i^{(N)}= c_i^{(N-1)}$. The proof for the direct part is complete.

\subsubsection{Converse}
\label{sec:converse}

If $p_{N-1}> N-1$, then from \Eq{eq:tmp222} and \Eq{eq:tmp333}, we
have $c_i^{(N)}\neq c_i^{(N-1)}$ at least for $j=N-1$, that is,
$i=p_{N-1}-(N-1)$. The proof is complete.

\subsection{Proof of Theorem~\ref{thm:minimal}}
The direct part of the theorem is trivial. To show the converse, let
$\tilde{\mn}\defeq (\tilde{n}_0,\tilde{n}_1,\ldots,\tilde{n}_N)$ and
$\tilde{\mn}'\defeq (\tilde{n}'_0,\tilde{n}'_1,\ldots,\tilde{n}'_{N'})$
be the two concerned minimal forms. In addition, we assume, without
loss of generality, that
\begin{align*}
  \tilde{n}_1&=\cdots=\tilde{n}_{i_1},\ldots,\tilde{n}_{i_{M-1}+1}=\cdots=\tilde{n}_{i_M}\\
  \tilde{n}'_1&=\cdots=\tilde{n}'_{i'_1},\ldots,\tilde{n}'_{i'_{M'-1}+1}=\cdots=\tilde{n}'_{i'_{M'}}
\end{align*}
with $i_M\leq N$ and $i'_{M'}\leq N'$. Now, let us define
$c_{0i}\defeq c_i - (1-i)$ with $c_i$ defined in \Eq{eq:ci3}. It can
be shown that $M$ intervals are non-trivial with
$\Abs{\Ical_{i_k}}\neq0$, $k=1,\ldots,M$. The values of $c_{0i}$'s are
in the following form
\begin{equation*}
  \overbrace{\ldots, \underbrace{\tilde{n}_{i_M},\ldots,\tilde{n}_{i_M}}_{i_M}}^{\Abs{\Ical_{i_M}}},\ \overbrace{\underbrace{\tilde{n}_{i_M}-1,\ldots,\tilde{n}_{i_M}-1}_{i_{M-1}}, \ldots, \underbrace{\tilde{n}_{i_{M-1}},\ldots,\tilde{n}_{i_{M-1}}}_{i_{M-1}}}^{\Abs{\Ical_{i_{M-1}}}},\ldots,\overbrace{\tilde{n}_2-1,\ldots,\tilde{n}_1+1,\tilde{n}_1}^{\Abs{\Ical_1}}.
\end{equation*}
Same arguments also apply to $\tilde{\mn}$ with $M'$ and $i'$, etc. It
is then not difficult to see that to have exactly the same
$c_{0i}$'s~(thus, same $c_{i}$'s), we must have $N=N'$ and
\begin{equation*}
  \tilde{n}_i = \tilde{n}'_i,\ \forall i=0,\ldots,N,
\end{equation*}%
that is, the same minimal form.

\section{Proof of Theorem~\ref{thm:recursive}}\label{sec:proof-thm-recursive}
\subsection{Sketch of the Proof}
\newcommand{\llra}{\Longleftrightarrow} To prove the theorem, we will
first show the following equivalence relations~:
\begin{align*} 
   (\R_1^{(N)}(k),\R_3^{(N)}(i,k)) &\stackrel{(a)}{\llra}   (\R_1^{(N)}(k),\R_2^{(N)}(i)),\quad \forall i,k;\\
   \R_3^{(N)}(i,k) &\stackrel{(b)}{\llra}   \R_3^{(N)}(N-1,k),\quad \forall i,k;\\
   (\R_1^{(N)}(k),\R_2^{(N)}(N-1)) &\stackrel{(c)}{\llra}   (\R_1^{(N)}(k),\R_2^{(N)}(i)\ \text{with ordered $\mn$});\\
   (\R_1^{(N)}(k),\R_2^{(N)}(i)\ \text{with ordered $\mn$}) &\stackrel{(d)}{\llra} (\R_1^{(N)}(k),\R_2^{(N)}(N-1)\ \text{with ordered and minimal $\mn$}).
\end{align*} 
\subsubsection{Equivalences $(a)$ and $(b)$} 
The direct parts of $(a)$, $(b)$, and $(d)$ are immediate since the
RHS are particular cases of the left hand side~(LHS).  To show the
reverse part of (a), we rewrite
\begin{align}
  d_{(n_0,\ldots,n_N)}(k) &= d_{(n_0-k,\ldots,n_N-k)}(0) \label{eq:tmp01}\\
  &= \min_{j\geq 0} d_{(n_0-k,\ldots,n_i-k)}(j) + d_{(j,n_{i+1}-k,\ldots,n_N-k)}(0)\label{eq:tmp02}\\
  &= \min_{j'\geq k} d_{(n_0,\ldots,n_i)}(j') +
  d_{(j',n_{i+1},\ldots,n_N)}(k)\label{eq:tmp03}
\end{align}
where $\R_1$ is used twice in \Eq{eq:tmp01} and \Eq{eq:tmp03}; $\R_2$
is used in \Eq{eq:tmp02}. As for (b), if $\R_3^{(N)}(N-1,k)$ holds, then
\begin{align}
  d_{(n_0,\ldots,n_N)}(k) &= \min_{j\geq k} d_{(n_0,\ldots,n_{N-1})}(j) + d_{(j,n_N)}(k)\label{eq:tmp04}\\
  &= \min_{j'\geq j\geq k} d_{(n_0,\ldots,n_{N-2})}(j')  + d_{(j',n_{N-1})}(j) + d_{(j,n_N)}(k) \label{eq:tmp05}\\ 
  &= \min_{j'\geq k} d_{(n_0,\ldots,n_{N-2})}(j')  + d_{(j',n_{N-1},n_N)}(k) \label{eq:tmp06}
\end{align}%
which proves $\R_3^{(N)}(N-2,k)$. By continuing the process, we can show
that $\R_3^{(N)}(i,k)$ is true for all $i$, provided $\R_3^{(N)}(N-1,k)$
holds. 
\subsubsection{Equivalences $(c)$ and $(d)$}
Through $(a)$ and $(b)$, one can verify that the LHS of $(c)$ is
equivalent to the RHS of $(a)$ of which the RHS of $(c)$ is a
particular case. Hence, the direct part of $(c)$ is shown. The
reverse part of (c) can be proved by induction on $N$. For $N=2$,
$\R_2^{(N)}(N-1)$ can be shown explicitly using the direct
characterization \Eq{eq:dk}. Now, assuming that $\R_2^{(N)}(N-1)$ for
non-ordered $\mn$, we would like to show that $\R_2^{N+1}(N)$ holds.
Let us write
\begin{align}
  \min_{j\geq 0} d_{(n_0,\ldots,n_{N})}(j) + d_{(j,n_{N+1})}(0)
  &= \min_{j\geq 0} d_{(\tilde{n}_0,\ldots,\tilde{n}_{i-1},\tilde{n}_{i+1},\ldots,\tilde{n}_{N+1})}(j)  + d_{(j,\tilde{n}_i)}(0) \label{eq:tmp001}\\
  &= \min_{k\geq j\geq 0} d_{(\tilde{n}_0,\ldots,\tilde{n}_{i-1},\tilde{n}_{i+1},\ldots,\tilde{n}_{N})}(k)  + d_{(k,\tilde{n}_{N+1})}(j) + d_{(j,\tilde{n}_i)}(0) \label{eq:tmp002}\\
  &= \min_{k\geq j'\geq 0} d_{(\tilde{n}_0,\ldots,\tilde{n}_{i-1},\tilde{n}_{i+1},\ldots,\tilde{n}_{N})}(k)  + d_{(k,\tilde{n}_{i})}(j') + d_{(j',\tilde{n}_{N+1})}(0) \label{eq:tmp003}\\
  &= \min_{j'\geq 0} d_{(\tilde{n}_0,\ldots,\tilde{n}_{N})}(j')  + d_{(j',\tilde{n}_{N+1})}(0) \label{eq:tmp003} \nonumber\\
  &=  d_{(n_0,\ldots,n_{N+1})}(0)\nonumber
\end{align}%
where the permutation invariance property is used in \Eq{eq:tmp001};
$\R_3^{(N)}(N-1,k)$ is used in \Eq{eq:tmp002} since we assume that
$\R_2^{(N)}(N-1)$ is trues; $\tilde{n}_i$ and $\tilde{n}_{N+1}$ can be
permuted according to $\R_2^{(2)}(1)$. Finally, we should prove the
reverse part of (d), \ie, 
\begin{equation}
  \label{eq:tmp004}
  d_{(\tilde{n}_0,\ldots,\tilde{n}_{N})}(0) = \min_{j\geq 0} d_{(\tilde{n}_0,\ldots,\tilde{n}_{N-1})}(j)  + j\tilde{n}_N 
\end{equation}%
provided that $\R_2^{(N)}(N-1)$ holds for minimal $\mn$.

If $\mn$ is not minimal, then showing (c) is equivalent to showing 
\begin{equation}
  \label{eq:tmp005}
    d_{(\tilde{n}_0,\ldots,\tilde{n}_{N^*})}(0) = \min_{j\geq 0} d_{(\tilde{n}_0,\ldots,\tilde{n}_{N^*})}(j)  + j\tilde{n}_N 
\end{equation}%
where $N^*$ is the order of $\mn$ with $\tilde{n}_{N^*+1}\leq
\tilde{n}_{N}$. Therefore, we should show that the minimum is achieved
with $j=0$. According the direct characterization~\Eq{eq:dk}, this is
true only when $\tilde{n}_N\geq c_1$. Let us rewrite $c_1$ as
\begin{align*}
  c_1 &= \left\lfloor\frac{\sum_{l=0}^{N^*}\tilde{n}_l - 1}{N^*} \right\rfloor \\
&= \left\lfloor\frac{N^*\tilde{n}_{N^*+1} + p_{N^*} - 1}{N^*} \right\rfloor.
\end{align*}%
Since $p_{N^*}\geq N^*$ is always true according to the reduction
theorem, we have $c_1\leq \tilde{n}_{N^*+1}\leq \tilde{n}_N$. The rest
of this section is devoted to proving that \Eq{eq:tmp004} holds for
minimal $\mn$.

\subsection{Minimal $\mn$}
Now, we restrict ourselves in the case of minimal and ordered $\mn$,
\ie, we would like to prove
\begin{equation}
  \label{eq:tmp006}
    d_{(\tilde{n}_0,\ldots,\tilde{n}_{N^*})}(0) = \min_{j\geq 0} d_{(\tilde{n}_0,\ldots,\tilde{n}_{N^*-1})}(j)  + j\tilde{n}_N.
\end{equation}%
Since 
\begin{align*}
c_{p_{N^*-1}} &= \tilde{n}_{N^*} + 1 -  p_{N^*-1} \\
      &\leq \tilde{n}_{N^*} + 1 - N^* \\
      &\leq \tilde{n}_{N^*},
\end{align*}%
the optimal $j$ is in the interval $\Ical_{N^*}\defeq
[1,p_{N^*-1}]$. Now, showing \Eq{eq:tmp006} is equivalent to showing 
\begin{equation*}
\sum_{i=1}^{p_{N^*-1}} 1-i+ \left\lfloor\frac{\sum_{l=0}^{N^*}\tilde{n}_l - i}{N^*} \right\rfloor = \min_{p_{N^*-1}\geq j \geq 0} \sum_{i=j+1}^{p_{N^*-1}} 1-i+ \left\lfloor\frac{\sum_{l=0}^{N^*-1}\tilde{n}_l - i}{N^*-1} + j\tilde{n}_{N^*}\right\rfloor 
\end{equation*}%
which, after some simple manipulations, is reduced to
\begin{equation}\label{eq:tmp007}
\sum_{i=1}^{p_{M}} \left( i - p_M + \left\lfloor\frac{i-1}{M+1} \right\rfloor \right) = \min_{k} \sum_{i=1}^{k} \left( i - p_M + \left\lfloor\frac{i-1}{M} \right\rfloor \right)
\end{equation}%
where we set $M\defeq N^*-1$ for simplicity of notation. Obviously,
the minimum of the RHS of \Eq{eq:tmp007} is achieved with such $k^*$
that 
\begin{align}
  k^* - p_M + \left\lfloor\frac{k^*-1}{M} \right\rfloor &\leq 0, \label{eq:tmp008-1}\\
\text{and}\  (k^*+1) - p_M + \left\lfloor\frac{k^*}{M} \right\rfloor &> 0. \label{eq:tmp008-2}
\end{align}%
Let us decompose $k^*$ as $k^*=aM+b$ with $b\in[1,M]$. Then, \Eq{eq:tmp008-1} becomes
\begin{equation}
  \label{eq:tmp009-1}
  aM+b-p_M+a \leq 0
\end{equation}%
which also implies that $aN+1-p_M+a\leq 0$ from which 
\begin{equation*}
  a = \left\lfloor \frac{p_M-1}{M+1} \right\rfloor.
\end{equation*}%
The form of $a$ suggests that $p_M$ can be decomposed as
\begin{equation}
  p_M = a(M+1) + \bar{b}. \label{eq:tmp009-2}
\end{equation}%
From \Eq{eq:tmp009-1} and \Eq{eq:tmp009-2}, we have $b\leq \bar{b}$ and
thus $b=\min\left\{M,\bar{b}\right\}$. With the form of optimal $k$
and some basic manipulations, we have finally
\begin{equation*}
  \sum_{i=1}^{p_{M}} \left( i - p_M + \left\lfloor\frac{i-1}{M+1} \right\rfloor \right) -  \sum_{i=1}^{k^*} \left( i - p_M + \left\lfloor\frac{i-1}{M} \right\rfloor \right)
 = 0
\end{equation*}%
which ends the proof.

\section{Proof of Theorem~\ref{thm:PF}}\label{sec:proof-thm-PF}
It can be proved by showing a stronger result~: the asymptotical pdf
of $\malpha({\RPPFtran} {\RPPF})$ in the high SNR regime is
identical to that of $\malpha(\transc{\RP}\, {\RP})$. We show it by
induction on $N$. For $N=1$, since $\umH_1 = \mH_1$, the result is
direct. Suppose that the theorem holds for $N$. Let us show that it
also holds for $N+1$. Note that
  $$\RPpPF=\umH_{N+1} \mP_N \RPPF = \umH_{N+1} \umD_N
  \transc{\umQ}_N \RPPF,$$
  from which we have
  \begin{align*}
    {\RPpPFtran}\, {\RPpPF} &\sim
    \Wcal_{n_0}(n_{N+1},\transc{(\umD_N \transc{\umQ}_N
      \RPPF)}(\umD_N \transc{\umQ}_N \RPPF)) \\
    &\sim \Wcal_{\underline{\nmin}}(n_{N+1},\mlambda(\transc{(\umD_N
      \transc{\umQ}_N \RPPF)}(\umD_N \transc{\umQ}_N \RPPF)))
  \end{align*}
  for a given $\RP$. Similarly, $\transc{\RP'}\, {\RP'} \sim
  \Wcal_{{\nmin}}(n_{N+1},\mlambda(\transc{\RP}\, \RP))$. In the
  high SNR regime, we can show that
  \begin{align*}
    \malpha(\transc{(\umD_N \transc{\umQ}_N \RPPF)}(\umD_N
    \transc{\umQ}_N \RPPF)) & = \malpha(\transc{(\transc{\umQ}_N
      \RPPF)}(\transc{\umQ}_N \RPPF)) \\
    &= \malpha({\RPPFtran}\,\RPPF)
  \end{align*}%
  where the first equality comes from
  lemma~\ref{lemma:invariance-asymp} and the second one holds because
  $$\transc{(\transc{\umQ}_N \RPPF)}(\transc{\umQ}_N \RPPF) =
  {\RPPFtran}\,\RPPF.$$
  Finally, since we suppose that the
  joint pdf of $\malpha({(\RPPFtran)}\,\RPPF)$ is the same as
  that of $\malpha(\transc{\RP}\,\RP)$, we can draw the same
  conclusion for $\malpha({(\RPpPFtran)}\,\RPpPF)$ and
  $\malpha(\transc{(\RP')}\,\RP')$.

\newpage

\newpage

\newpage

\end{document}